\documentclass[aps,pra,twocolumn,superscriptaddress]{revtex4-1}
\usepackage{csquotes}
\usepackage{dcolumn}
\usepackage{bm}
\usepackage{relsize}		
\usepackage{url}	
\usepackage{physics}
\usepackage{amsfonts}
\usepackage{float}
\usepackage{bbold}
\usepackage{amssymb}
\usepackage{amsmath}
\usepackage{bbold}
\usepackage{graphicx}
\usepackage{hyperref}
\graphicspath{{pics/}}
\usepackage{ulem}

\usepackage[usenames,dvipsnames]{xcolor}
 \definecolor{david}{rgb}{0.7,0,0.9}

\begin{document}

\title{Inaccessible entanglement in symmetry protected topological phases}
\author{Caroline de Groot}
\affiliation{
Max-Planck-Institute of Quantum Optics, Hans-Kopfermann-Str.~1, 85748 Garching,
 Germany  }
 \affiliation{Munich Center for Quantum Science and Technology, Schellingstr.~4, 80799 M\"unchen, Germany}
\author{David T. Stephen} 
\affiliation{
Max-Planck-Institute of Quantum Optics, Hans-Kopfermann-Str.~1, 85748 Garching,
 Germany  }
 \affiliation{Munich Center for Quantum Science and Technology, Schellingstr.~4, 80799 M\"unchen, Germany}
\author{Andras Molnar}
\affiliation{
Max-Planck-Institute of Quantum Optics, Hans-Kopfermann-Str.~1, 85748 Garching,
 Germany  }
 \affiliation{Munich Center for Quantum Science and Technology, Schellingstr.~4, 80799 M\"unchen, Germany}
\affiliation{Instituto de Ciencias Matem\'aticas, Campus Cantoblanco UAM, C/ Nicol\'as Cabrera, 13-15, 28049 Madrid, Spain}
\affiliation{Dpto.\ An\'alisis Matem\'atico y Matem\'atica Aplicada, Universidad Complutense de Madrid, 28040 Madrid, Spain}
 \author{Norbert Schuch} 
\affiliation{
Max-Planck-Institute of Quantum Optics, Hans-Kopfermann-Str.~1, 85748 Garching,
 Germany  }
 \affiliation{Munich Center for Quantum Science and Technology, Schellingstr.~4, 80799 M\"unchen, Germany}

\date{\today}

\begin{abstract}
We study the entanglement structure of symmetry-protected topological (SPT) phases from an operational point of view by considering entanglement distillation in the presence of symmetries. We demonstrate that non-trivial SPT phases in one-dimension necessarily contain some entanglement which is inaccessible if the symmetry is enforced. More precisely, we consider the setting of local operations and classical communication (LOCC) where the local operations commute with a global onsite symmetry group $G$, which we call $G$-LOCC, and we define the inaccessible entanglement $E_{inacc}$ as the entanglement that cannot be used for distillation under $G$-LOCC. We derive a tight bound on $E_{inacc}$ which demonstrates a direct relation between inaccessible entanglement and the SPT phase, namely $\log(D_\omega^2)  \leq E_{inacc} \leq \log(|G|)$, where $D_\omega$ is the topologically protected edge mode degeneracy of the SPT phase $\omega$ with symmetry $G$. For particular phases such as the Haldane phase, $D_\omega = \sqrt{|G|}$ so the bound becomes an equality. We numerically investigate the distribution of states throughout the bound, and show that typically the region near the upper bound is highly populated, and also determine the nature of those states lying on the upper and lower bounds. We then discuss the relation of $E_{inacc}$ to string order parameters, and also the extent to which it can be used to distinguish different SPT phases of matter.

\end{abstract}

\maketitle

\section{\label{intro}Introduction}

Entanglement is the essential resource which allows for tasks beyond the restrictions of local operations and classical communication (LOCC), that are impossible to implement classically, such as quantum teleportation, dense coding and secure cryptography \cite{bennett1993teleporting, ekert1991quantum,bennett1992communication}. Under LOCC, the quantification of entanglement leads to the von Neumann entropy as the determining quantity governing state conversion \cite{nielsen2002quantum}. By imposing additional restrictions to LOCC, the non-local resources that emerge, as well as the tasks they allow, are reshaped. In a seminal work, Wiseman and Vaccaro found that the entanglement of indistinguishable  particles subject to a super-selection rule (SSR) fixing the total particle number is reduced to the average of the von Neumann entropy of spatial modes over all particle number sectors for a mode \cite{wiseman2003entanglement}. This pioneering observation has now led to a plethora of interdisciplinary works in the study of SSR in the quantum information community \cite{PhysRevA.100.022324, barghathi2019operationally,klich2008scaling,barghathi2018renyi,goldstein2018symmetry,wiseman2003entanglement,xavier2018equipartition, vaccaro2008tradeoff}. SSR refers to phenomena where particular state transitions are forbidden in the Hilbert space due to some physical rule, such as disallowing a superposition across different electronic charge sectors, which would violate the inherent parity symmetry of fermions \cite{wick1970superselection}, or restrictions due to particle number conservation \cite{PhysRevA.70.042310}. However, the interplay between entanglement and SSR in physical systems has yet to be fully explored. Bridging this gap firstly necessitates modifying the rules of LOCC to include the SSR which enforce further structure on the partial order of states due to a redefinition of the majorisation criterion \cite{PhysRevA.70.042310}. SSR were shown to even produce additional non-local resources, \textit{e.g.}\ to permit perfect quantum data hiding protocols \cite{verstraete2003quantum}. This motivated exploring the entanglement under restricted settings in various contexts, such as the symmetry-resolved entanglement \cite{2002.04620,fraenkel2019symmetry, goldstein2018symmetry,2002.04367,cornfeld2019entanglement}, as well as conservation laws in fermionic systems with parity symmetry and ultracold atoms with particle number conservation, for both critical and gapped systems \cite{bartlett2003entanglement,1911.09588,laflorencie2014spin}. In the context of lattice gauge theories, which impose a local gauge symmetry, it was shown that entanglement distillation must be adjusted to account for additional restrictions \cite{ghosh2015entanglement,PhysRevLett.117.131602,soni2016aspects}. Here we are concerned with the impact of enforcing a global symmetry, leading us to consider symmetry protected topological (SPT) phases \cite{chen2010local,prevWork3,wen2017colloquium,pollmann2010,else2014classifying,schuch2011classifying,chen2011classification}. \\

The motivation of this work is to apply insights about entanglement under SSR from the quantum information perspective to SPT phases. This as yet unexplored connection is natural to make, as topological phases describe order beyond the Landau description of symmetry breaking and local order parameters, and are instead determined by global patterns of entanglement \cite{wen2019choreographed,prevWork2,duivenvoorden2017entanglement}. In particular, topological phases are classified through non-local order. The many-body entanglement picture of topological order sorts phases into two classes; topological phases have long range entanglement (i.e. have topological order), while others have short range entanglement (i.e. have SPT order). We are concerned with the latter, which only host topological properties in the presence of symmetry, such as topological insulators and the Haldane chain \cite{wen2017colloquium,prevWork3, wen2019choreographed}. The development of quantities which detect SPT order is of particular interest in improving intuitions about these phases \cite{verresen2017one,wen2002quantum}, such as the string order parameter and SPT-entanglement which characterise the phase in terms of the global entanglement structure \cite{iman,pollmann2012detection,PhysRevB.40.4709, haegeman2012order,stephen2019detecting}. \\

In this work, we study the entanglement of SPT ordered systems under local operations commuting with a global onsite symmetry $G$, which we call $G$-LOCC. This setting is naturally suggestive of SPT phases which are only defined with respect to the symmetry group $G$ protecting the order. This classification breaks down when considering scenarios with no symmetry. In $G$-LOCC, the amount of accessible entanglement is naturally reduced as compared to LOCC. We therefore show that, in the presence of non-trivial SPT order under $G$, there is always some entanglement that is inaccessible under $G$-LOCC. We give this quantity operational meaning by showing that it is consistent with entanglement distillation under $G$-LOCC. In more detail, we study the inaccessible entanglement $E_{inacc}$ of SPT phases in one-dimension (1D), which leads to the following bound
\begin{equation}
\log(D_\omega^2)  \leq E_{inacc} \leq \log(|G|),
\end{equation}
where $D_\omega$ is the topologically protected part of the edge mode degeneracy and the dimension of the projective irreducible representation defining the SPT phase $\omega$. We show that maximally non-commutative (MNC) phases, such as the Haldane or cluster phase for the symmetry $\mathbb{Z}_2 \times \mathbb{Z}_2$, saturate the upper bound on inaccessible entanglement. This is connected to the fact that MNC phases maximise the number of topologically protected degenerate edge modes. Importantly, trivial SPT phases always have zero as the lower bound, while non-trivial phases have a non-zero lower bound. \\

In order to obtain a better understanding of the typical behaviour of the inaccessible entanglement, we numerically study particular random distributions of states, using matrix product states (MPS). We demonstrate that the bound is tight by interpolating through a path of states from the lower bound to the upper bound. Intriguingly, we find that both trivial and non-trivial SPT systems typically have near maximal inaccessible entanglement. However, the irrep probabilities, which are the weights of the state on the individual symmetry sectors, have different structure, and can thus be used to distinguish these phases. We study the implication of effective reduced symmetries in a non-trivial SPT system, where we interpolate a state with a symmetry $G$ to effectively a lower symmetry $H \subset G$. The interpolation causes the inaccessible entanglement and the irrep probabilities to lose the structure due to $G$ and the final state displays only structure due to $H$. On the other hand, we also study the effect of reduced $G$-LOCC, so that the local operations commute with some $H \subset G$ where $G$ is still the symmetry protecting the SPT phase, as opposed to the effective reduced symmetry scheme. We argue that the accessible entanglement depends on the dimension of the subspace of the MPS which transforms trivially under the symmetry, previously termed the junk subspace \cite{prevWork1}. We can increase the dimension of the junk subspace for a given bond dimension by giving an uneven irrep structure to the projective representation, which causes a greater spread in the distribution of inaccessible entanglement. Finally, we discuss how our results apply to two-dimensional subsystem SPT phases \cite{you2018subsystem,stephen2019subsystem}, and also how they relate to the characterization of states as resources for measurement-based quantum computation \cite{raussendorf2003measurement}. \\

The manuscript is structured as follows. In Section~\ref{section2} we provide the necessary background on SPT order and the description with tensor networks. In Section \ref{inaccprotocol}, we formalise the definition of inaccessible entanglement via entanglement distillation with $G$-LOCC. Then, in Section \ref{bound1}, we present our proofs of the bound on inaccessible entanglement for all SPT phases under finite Abelian symmetry groups, and the connection to string order parameters. Next, in Section \ref{investigation} we confirm that the bound is tight, and examine properties of SPT phases living in particular regions of the bound by performing interpolations in the MPS tensor. In Section \ref{discuss} we discuss the implications on computational power, subsystem SPT order, and restricting operations to a subgroup of $G$. Finally, in Section \ref{conclusion} we conclude and suggest future directions enabled by this work.

\section{Introduction to SPT phases with tensor networks\label{section2}}

In this Section we review the necessary frameworks to discuss SPT phases with tensor networks. Throughout this work, we will restrict to finite Abelian symmetries. This is not a severe restriction, as many physically relevant SPT phases protected by other groups remain non-trivial even when a finite Abelian subgroup of the symmetry is enforced \cite{stephen2017computational}. One example is the Haldane phase, which may be defined as the non-trivial SPT phase protected by $SO(3)$ symmetry, but is effectively protected by a $\mathbb{Z}_2 \times \mathbb{Z}_2$ subgroup. The non-trivial phase protected by the latter symmetry is also called the cluster phase, as it contains the 1D cluster state \cite{raussendorf2003measurement}. Being the simplest non-trivial SPT phase, we will often refer to this example throughout the paper.

\subsection{Introduction to tensor networks}

A powerful tool to describe topological phases is the tensor network formalism, in which the wave function is decomposed into a network of local tensors $A^i$ associated to each site. Tensor networks are ideal candidates for capturing states with symmetries and/or topological order, as they allow us to understand global properties of a states in terms of symmetries of the local tensors, which in turn describe the entanglement structure of the state \cite{schuch2011classifying,chen2011classification}.\\
 
In this paper we consider the one-dimensional tensor network states called matrix product states (MPS). With this ansatz one can efficiently approximate states obeying an area law in entanglement entropy, such as ground states of gapped, local Hamiltonians \cite{hastings2007area,verstraete2006matrix}.
 An MPS is defined by a single rank-three tensor $A$ as
\begin{equation} \label{mpstens}
    \ket{\Psi[A]}= \sum_{i_1, \dots, i_N} \Tr(A^{i_1} \dots A^{i_N}) \ket{i_1 \dots i_N},
\end{equation}
where $A^i$ are matrices such that $A=\sum_i A^i \otimes \ket{i}$. From now on, we only consider injective MPS, which are defined as those for which the transfer matrix $T = \sum_i A^i \otimes (A^i)^\dagger$ has a unique eigenvalue of largest magnitude. Physically, this condition holds when the MPS is the unique ground state of its parent Hamiltonian, which is indeed the case for states in SPT phases without symmetry-breaking.\\

\begin{figure}
    \centering
    \includegraphics[width=\linewidth]{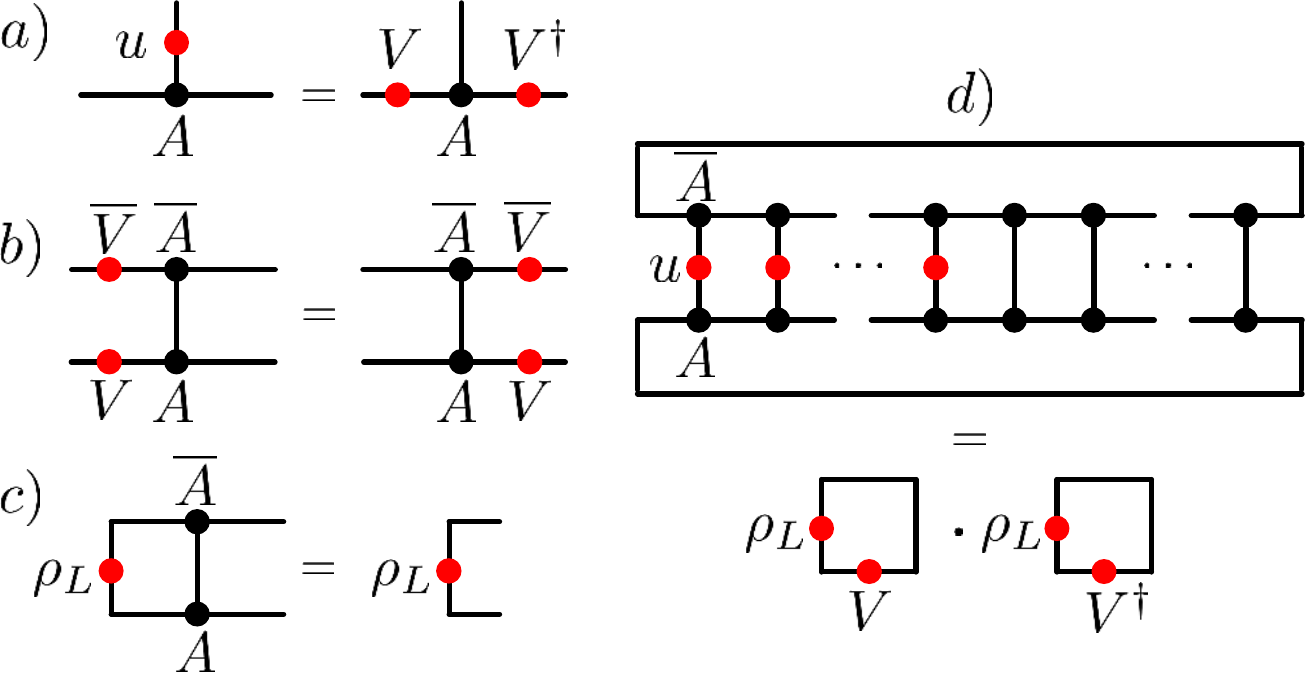}
    \caption{In this figure, $u$ and $V$ have a dependence on $g$ which is omitted. 1a) Graphical representation of Eq.~\ref{fundtheorem}. 1b) Symmetries of the transfer operator $T$. 1c) Definition of the left-fixed point $\rho_L$. 1d) Graphical calculation of the trace part of the irrep probabilities leading to \eqref{probs}. }
    \label{fig:tensornetworks}
\end{figure}

Injective MPS satisfy an important property, often called the fundamental theorem of MPS, which states that any global onsite symmetry,  $u(g)^{\otimes N} \ket{\Psi[A]} = \ket{\Psi[A]}$, implies that the local tensor $A$ transforms under the on-site symmetry action as \begin{equation}\label{fundtheorem}
    \sum_j u(g)_{ij} A^j= e^{i\phi(g)}V(g) A^i V(g)^{\dagger},
\end{equation} where $u(g)$ is some group representation and $V(g)$ is a projective representation, as pictured in Fig.~\ref{fig:tensornetworks} a) \cite{perez2006matrix}. We take $A$ to be in canonical form so that $V(g)$ is unitary. The projective representation differs from a linear representation as it obeys the relation \begin{equation}
     V(g)V(h) = \omega(g,h) V(gh)
\end{equation} with a phase $\omega$ called a cocycle. The fundamental theorem elucidates that the projective symmetry action of $V(g)$ occurs on the virtual level of each site tensor $A$, such that the entanglement structure and topological properties of the global state are encoded locally. As such the properties of $V(g)$ play an important role in tensor networks.\\

The transfer operator $T$, which inherits the symmetries of $A$ as shown in Fig.~\ref{fig:tensornetworks} b), is an important object in coming calculations, so we establish some notation here. Let $\ket{R}$ and $\bra{L}$ denote the right and left eigenvectors of $T$ corresponding to the largest eigenvalue, which we can set to be equal to 1 with normalisation, such that $T \ket{R} = \ket{R}$ and $\bra{L} T = \bra{L}$. For the purposes of this work, it is often convenient to express $T$ as a quantum channel $\mathbb{T}$ defined as $\mathbb{T}(\rho) = \sum_i A^i \rho A^{i \dag}$. Then, $\ket{R}$ and $\bra{L}$ correspond to the fixed-points $\rho_R$ and $\rho_L$ of $\mathbb{T}$ and $\mathbb{T}^\dagger(\rho)=\sum_i A^{i \dag} \rho A^{i}$, respectively, see Fig~\ref{fig:tensornetworks} c).

\subsection{Classification of SPT phases}

SPT order is defined by equivalence classes of short-range entangled states, \textit{i.e.} families of states which are connected by a finite-depth quantum circuit respecting a particular symmetry, which corresponds to an adiabatic path of gapped, local Hamiltonians \cite{chen2010local,prevWork3,wen2017colloquium}. As such, this order is called ``topologically trivial'', but nevertheless intrinsically non-local properties prevail in presence of a symmetry, such as topologically protected edge mode degeneracy and string order \cite{den1989preroughening,iman,pollmann2012detection}; this follows from the non-trivial symmetry action of the projective representation $V(g)$ on the edges \cite{pollmann2010,else2014classifying}.  \\

It has been shown that 1D SPT phases are classified by the second cohomology group $H^2(G,U(1))$ which is formed by the equivalence classes of cocycles $\omega$ up to coboundaries of a particular projective representation $V(g)$ \cite{pollmann2010,schuch2011classifying,chen2011classification}. Hence, SPT order is intimately linked to the projective representation $V(g)$, and remarkably demonstrates that global patterns of entanglement appear via a local mechanism, which emphasises the natural suitability of local tensor descriptions.  \\

In the following we focus on bosonic SPT order at zero temperature. For completeness we add that, generally, bosons in $d$-dimensional SPT phases of matter are classified by the $d+1$ group cohomology $H^{d+1}(G,U(1))$ \cite{prevWork3}, although models beyond group cohomology have also been studied in $3D$ and $4D$ \cite{wen2017colloquium,1912.05565}.

\subsection{Projective irreps of Abelian groups}
We now introduce several properties about projective irreducible representations of finite Abelian groups that will be relevant in later calculations.\\

Let us introduce the term $\omega$-irrep to refer to an irreducible representation with cocycle $\omega$. Firstly, a general projective representation $V(g)$ of a finite group $G$ with cocycle $\omega$ can be  written as a direct sum over $\omega$-irreps $V_a(g)$
\begin{equation}
    \label{projectiverep} V(g) = \bigoplus_{a} \left( \mathbb{1}_{n_a} \otimes V_a(g) \right),
\end{equation}
where $n_a$ is the multiplicity for a given $\omega$-irrep $a$ \cite{karpilovsky1994group}. The second property we make use of is the fact that any two $\omega$-irreps of a finite Abelian group are projectively equivalent. Namely, if we fix $\widetilde{V}(g)$ to be an arbitrary reference $\omega$-irrep, every other $\omega$-irrep $V_a(g)$ can be related to it by \begin{equation}\label{projequivalence}
    V_a(g) = \mu_{a}(g) U_a \widetilde{V}(g) U_a^\dag,
\end{equation} where $\mu_{a}(g)$ can be thought of as a phase factor, as it is a linear character, and $U_a$ is some unitary \cite{backhouse1972projective}. 
This allows us to simplify the general form of $V(g)$ in Eq.~\eqref{projectiverep}, which is a direct sum over the $\omega$-irreps containing multiple tensor products, to a single tensor product when $G$ is finite Abelian. Namely, 
\begin{equation} \label{projgeneral} V(g) = \left(\bigoplus_a \mu_a(g)\mathbb{1}_{n_a}\right) \otimes \widetilde{V}(g), \end{equation} 
where we have removed the unitaries $U_a$ through an appropriate choice of basis. This illustrates that, for finite Abelian groups, all $\omega$-irreps for a given $\omega$ have the same dimension, which we call $D_\omega$.\\

To calculate $D_\omega$, we need to introduce the notion of the projective centre group $k_\omega$. Given a cocycle $\omega$ under the group $G$,  $k_\omega$ is defined by
\begin{equation} \label{projcentre}
    k_\omega = \{ s \in G \hspace{1ex}| \hspace{1ex}\omega(g,s)= \omega(s,g)\hspace{1ex} \forall g \in G \}.
\end{equation}
Then, $D_\omega$ is determined by the following equation \cite{berkovich2018yakov},
\begin{equation} \label{edgemodes}
    D_\omega = \sqrt{|G|/|k_\omega|}.
\end{equation} 
For example, the Haldane phase has projective irreps given by the 2-dimensional Pauli operators. Conversely, for non-Abelian groups the irreps may have different dimensions, an example of which is $S_4$, the symmetric group on four elements, which has 2- and 4-dimensional projective irreps.\\

Finally, we note that for any $\omega$-irrep $V_\omega (g)$, the following condition about the trace holds \cite{karpilovsky1994group},
\begin{equation}\label{theorem}
    \Tr(V_\omega(g)) =
    \begin{cases}
      D_\omega, & \text{if}\ g \in k_\omega \\
      0 & else.
    \end{cases}
 \end{equation}
This will be helpful for calculations later on. 

\subsubsection{Maximal non-commutativity}
SPT phases of special interest are the so-called maximally non-commutative (MNC) phases. These phases act as universal resources for MBQC \cite{raussendorf2003measurement,miyake2010quantum,prevWork1,miller2015resource,stephen2017computational}. A cocycle $\omega$, along with the corresponding SPT phase, is called MNC if $k_\omega = \{ e \}$, meaning it has a trivial projective centre. For such phases, $D_\omega$ takes the maximum value of $\sqrt{|G|}$. Since $D_\omega$ is equal to the topologically protected edge degeneracy of SPT phases \cite{pollmann2010}, MNC phases therefore have the maximum edge degeneracy for a given group $G$. From now on, we will suppress the index $\omega$ in $k_\omega$ for notational simplicity.\\

For each MNC cocycle $\omega$ there is only one $\omega$-irrep up to unitary equivalence \cite{berkovich2018yakov}, such that Eq.~\eqref{projgeneral} becomes
\begin{equation} \label{projrepmnc}
V(g) = \mathbb{1} \otimes \widetilde{V}(g).
\end{equation}
The left part of the tensor product, which transforms trivially under symmetry, is referred to as the junk subspace. The size of this space determines the bond dimension and hence bounds the entanglement entropy of the state. We will also refer to the trivially transforming part of the MPS tensor in non-MNC phases as the junk subspace, although strictly there are multiple subspaces, corresponding to the existence of multiple projective irreducible representations. \\
 
We note that some symmetry groups host multiple MNC phases; such phases can't be distinguished by their edge mode degeneracy, as the $\omega$-irreps all have the same dimension. Order parameters as in Refs. \cite{iman,pollmann2012detection,haegeman2012order,PhysRevB.40.4709}, however, can distinguish between them by microscopically probing the specific non-local action on the virtual level of the tensor network, which we discuss further in Section \ref{stringorder}. \\

\section{ \label{inaccprotocol} Accessible entanglement distillation}

How much entanglement can be extracted from a state obeying a symmetry $G$ under LOCC respecting the same symmetry? In this Section we define the accessible entanglement in the presence of $G$-symmetric LOCC ($G$-LOCC) and derive an expression through multi-shot entanglement distillation which is consistent with previous works \textit{e.g.} Refs. \cite{wiseman2003entanglement,klich2008scaling,barghathi2018renyi,goldstein2018symmetry,barghathi2019operationally}. We will first define $G$-LOCC and motivate why we might expect this to effect the entanglement which is accessible under these operations.\\

We define $G$-LOCC by restricting LOCC protocols to positive-operator valued measure (POVM) sets $\{M_\alpha^\dagger M_\alpha\}$ that commute with the global symmetry $U(g) = u(g)^{\otimes N}$ such that $[U(g),M_\alpha]=0$ $\forall g \in G$, where $\alpha$ are the measurement outcomes. Note that the $M_\alpha$ are not necessarily orthogonal measurements. The accessible entanglement we will motivate that exists under such POVM sets is given by \begin{equation} \label{accequation}
    E_{acc}(\rho) = \sum_\alpha p_\alpha E(\rho_\alpha),
\end{equation} where $p_\alpha = \Tr \left( M_\alpha \rho M_\alpha^\dag \right)$ are measurement probabilities corresponding to the post-measurement state $\rho_\alpha = M_\alpha \rho M_\alpha^\dag $ up to normalisation and the entanglement entropy $E \left( \rho_{AB} \right) $ of a bipartite state $\rho_{AB}$ is given by the von Neumann entropy $ S \left(\rho_A \right)$, where $\rho_A = \Tr_B\left( \rho_{AB} \right)$ is the reduced density operator \cite{nielsen2002quantum}. We will now argue that the entanglement entropy overestimates the physical entanglement $E_{acc}$ due to the structure imposed by symmetry. This line of thought leads naturally into the entanglement distillation protocol of the next Section, in which we will derive Eq.~\eqref{accequation}. \\

A simple argument considers that local operations may not mix symmetry sectors, and that local operations can only decrease entanglement. As the state $\rho$ is symmetric under the symmetry $G$ which imposes the $G$-LOCC, enforcing $[u(g)^{\otimes N },\rho ]=0$, the reduced density $\rho^A$ also has the symmetry $[u(g)^{\otimes N_A },\rho^A ]=0$. In the following we will make use of a particular POVM set to directly measure the symmetry and capture properties of the SPT phase under $G$-LOCC. We utilise the projective measurement $\Pi_\alpha$ onto the irrep sectors given by \begin{equation}\label{symproject}
    \Pi_\alpha = \frac{1}{|G|} \sum_g \chi_\alpha \left( g\right) u(g)^{\otimes N_A},
\end{equation} where $\chi_\alpha(g)$ is the character corresponding to irrep $\alpha$ and $N_A$ are the sites in the $A$-subsystem (see also \cite{iman}). Due to the symmetry, the reduced state can be written as $\rho^A = \bigoplus_\alpha p_\alpha \rho_\alpha^A$, where $p_\alpha$ are the measurement probabilities defined above for the POVM operators $M_\alpha = \Pi_\alpha$, and where $\rho_\alpha^A = \Pi_\alpha \rho^A \Pi_\alpha^\dag /p_\alpha$. \\

All allowed operations commute with the symmetry and hence operate only within single sectors $\alpha$ which can be written as $\rho_{\alpha,out} = \Lambda_\alpha (\rho_{in})$ with the map $\Lambda_\alpha( \cdot) = M_\alpha ( \cdot ) M^\dag_\alpha$, and $M_\alpha$ defining an element of a POVM set obeying the symmetry and acting on a single sector. These operations are local by definition. Since operations which mix the symmetry sectors of states are not supported, any entanglement between different sectors is erased by $G$-LOCC. Hence, only the entanglement within individual symmetry sectors contributes. The accessible entanglement defined in Eq.~\eqref{accequation} is therefore the sum over the entanglement within each symmetry sector $\alpha$. Clearly the entanglement in each sector $E(\rho_\alpha)$ is less than the total entanglement, but so is the weighted sum of $E(\rho_\alpha)$ over $\alpha$. As this defines the accessible entanglement, it is therefore also smaller or equal to the entanglement, \begin{equation}
  E_{acc} \leq E.
\end{equation} Hence $G$-LOCC leads to a reduced accessible entanglement. Note that this expression ensures that product states have $E_{acc} = 0$. The considerations above naturally lead to an operational definition of the accessible entanglement. 

\subsection{Entanglement distillation under $G$-LOCC}

We now present how entanglement distillation under $G$-LOCC leads us to the accessible entanglement in Eq.~\eqref{accequation}. Entanglement distillation is the process of distilling Bell pairs from asymptotically many copies of a state under LOCC, and is consistent with a partial order on states through the majorisation criterion \cite{nielsen2002quantum,wilde2013quantum}. This leads to the equivalence of the conversion ratio $M/N$, between $N$ copies of a pure state $\ket{\Psi}$ to $M$ copies of the Bell state in the asymptotic limit $N \rightarrow \infty $, to $E$ the entanglement entropy \cite{bennett1996concentrating}. We now consider entanglement distillation under $G$-LOCC.\\

We measure the POVM set $\{\Pi_\alpha^\dag\Pi_\alpha\}$ with $\Pi_\alpha$  as defined in Eq.~\eqref{symproject} the projector onto each irrep $\alpha$ of the symmetry $G$ imposing the $G$-LOCC, such that on average we get every channel containing the normalised output state $\rho_\alpha$ with a probability $p_\alpha$ in the asymptotic limit. This action splits the state into the symmetry sectors which impose the super-selection rule, so that operations are restricted to within these subspaces. This makes it possible to act with LOCC within each subspace. Then the entanglement entropy of the output state at each branch $\alpha$ computes the number of Bell pairs distilled from the $\alpha$-channel. Clearly, the total entanglement extracted from this procedure is the weighted sum of the entanglement from each branch, giving the expression in Eq.~\eqref{accequation}. \\

Furthermore, as the LOCC can be defined free of SSR within each sector, each branch contains an optimal distillation within each symmetry sector  such that the full distillation is optimal \cite{nielsen2002quantum}, \textit{i.e.} the maximal possible ratio $M/N$ of $M$ Bell pairs produced from $N$ copies of $\ket{\Psi_\alpha}$ is achieved. This concludes the description of entanglement distillation under $G$-LOCC, which employs the full rigour of distillation under LOCC, and as such is of itself complete.

\subsection{Symmetry measurement}

In order to carry out calculations later on, we rewrite the irrep probabilities in tensor network notation through some simple manipulations which we sketch here. We consider a system on a ring with $N$ sites in the limit of the number of sites going to infinity which allows us to use properties of the fixed point, such that bipartitioning cuts the ring in half and results in two boundaries. Using $p_\alpha = \Tr \left( \Pi_\alpha \rho \right)$, and inserting the form of the symmetric projection operator from Eq.~\eqref{symproject}, we should recall that $\Pi_\alpha$ acts only on the $A$-subsystem, so the rest is traced over, leaving the fixed point action on the remaining half. \\

By using the fact that the symmetry action in $\Pi_\alpha$ acts locally, we can pull the sum out and by engaging the fundamental theorem, Eq.~(\ref{fundtheorem}), we find an effective action of the symmetry group on the virtual system. Due to the cross-cancellation of all $V(g),V(g)^\dag$ except for the action at the edges, the rest of the expression becomes
\begin{equation} \label{probs2}
     p_\alpha = \frac{1}{|G|} \sum_g \chi_\alpha(g) \langle L| (\mathbb{1}\otimes V(g)) T^{N_A} (\mathbb{1}\otimes V(g)^\dag)|R\rangle.
\end{equation}
We now use that $T^{N_A} \to \ket{R}\bra{L}$ as $N_A\to\infty$, and for simplicity we switch from the language of vectors $\{\ket{L},\ket{R} \}$ to matrices $\{\rho_L,\rho_R \}$ which are the same object, the fixed point of the MPS, considered either as vectors or as matrices. Then the remaining expression is 
\begin{equation} \label{probs}p_\alpha = \frac{1}{|G|} \sum_g  {\chi_\alpha(g)} \Tr(\rho_L \ V(g)) \Tr(\rho_L V(g)^\dag), \end{equation}
using right-canonical form such that $\rho_R$ is the identity \cite{schollwock}. The part of the summand involving the trace in the calculations leading to Eq.~\eqref{probs} is shown pictorially in Fig.~\ref{fig:tensornetworks}.

\subsection{\label{definacc}Inaccessible entanglement}

Having defined the accessible entanglement by distillation under $G$-LOCC earlier, we now conversely define the inaccessible entanglement as the amount of entanglement that the $G$-LOCC prevents access to:
\begin{equation}\label{inacc_def}
    E_{inacc}= E - E_{acc}.
\end{equation}
While the accessible entanglement captures the physical entanglement of a system with SPT order, inaccessible entanglement measures the entanglement protected by the presence of symmetry and stored in the state. We show below that the inaccessible entanglement can be written as
\begin{equation} \label{inacc}
    E_{inacc} \left( \rho \right)= -\sum_\alpha\ p_\alpha \log \left( p_\alpha \right),
\end{equation} where $p_\alpha$ are the irrep probabilities defined above. In the following we will use $\log$ to signify log base-2. Note that Eq.~\eqref{inacc} is the entropy of the probability distribution corresponding to the irrep weights of the symmetry group $G$ on the state, $E_{inacc} = S(\{p_\alpha\})$. \\

Recall that the reduced density $\rho^A$ has the symmetry $[u(g)^{\otimes N_A },\rho^A ]=0$, and therefore can be written as $\rho^A = \bigoplus_\alpha p_\alpha \rho_\alpha^A$ where $\rho_\alpha = \Pi_\alpha \rho^A /p_\alpha$. Therefore the entanglement entropy of the state is
\begin{align}
    E(\rho) = -\sum_\alpha p_\alpha \rho_\alpha^A \log(p_\alpha \rho_\alpha^A) = \sum_\alpha p_\alpha E(\rho_\alpha) - p_\alpha \log p_\alpha.
\end{align} Then Eq.~\eqref{inacc} follows since the first term in the second equality is the accessible entanglement, and subtracting this gives the relation for inaccessible entanglement. Notice that since $E \geq E_{inacc}$, the presence of entanglement is a necessary condition for inaccessible entanglement. \\

Given the irrep probabilities in Eq.~\ref{probs}, we now have a tensor network formulation of $E_{inacc}$. In the following, we explore how different classes of SPT phases behave with regards to inaccessible entanglement.

\section{\label{bound1}Bounds for the inaccessible entanglement}

In this Section we derive bounds on the entanglement that is inaccessible due to SPT order. It is trivial to deduce that $E_{inacc} = 0$ for a product state (which has no entanglement) or a state with no symmetry (which has no SSR). We show, however, a non-zero lower bound in the presence of symmetries for states with non-trivial SPT order. This bound, together with the upper bound, reads as
\begin{equation} 
  \log(\frac{|G|}{|k|})  \leq E_{inacc} \leq \log(|G|),
\end{equation} 
where $G$ is the symmetry group protecting the topological order and $k$ is the projective centre of $G$ associated to the cocycle characterizing the SPT phase (defined in Eq.\ \eqref{projcentre}). We thereby develop a prescription for all SPT phases under finite Abelian groups by showing that non-trivial SPT phases remarkably incur a non-zero lower bound on the inaccessible entanglement. 
  
\subsection{Maximal inaccessible entanglement in the MNC phase \label{maxinacc}}

We first consider the simpler case of MNC phases, which obey a strong condition on inaccessible entanglement where the lower bound meets the upper bound \begin{equation} \label{mnceq}
    E_{inacc} = \log(|G|),
\end{equation} given the symmetry $G$. MNC phases therefore maximise the inaccessible entanglement. The irrep probabilities are $p_\alpha = \frac{1}{|G|}$ $ \forall \alpha$, which produces maximal entropy $S(\{p_\alpha\})$, and hence Eq.~\eqref{mnceq}. We first give an example to illustrate this result. \\

The Haldane or cluster phase is a canonical example of the MNC phase, which is protected by the symmetry $G = \mathbb{Z}_2 \times \mathbb{Z}_2$ for which the cohomology group is known to be $H^2(G,U(1)) = \mathbb{Z}_2$, indicating one trivial and one SPT phase \cite{prevWork2,prevWork1}. One valid symmetry representation is given by $u(g)= \bigoplus_\alpha \chi_\alpha(g)$, where $\chi_\alpha$ is the character of the irrep $\alpha$, and the projective representation in the MNC phase is given by the Pauli operators including the identity, $V(g) = \sigma(g)$. States satisfying this symmetry have an MNC cocycle as all the Pauli operators mutually anti-commute, so the projective centre $k$ contains only the identity. Since the Pauli matrices are traceless, $\Tr(\sigma(g))= 0$ $\forall g \in G$, the only contribution to $p_\alpha$ in Eq.~\eqref{probs} is from the identity element and hence $p_\alpha = \frac{1}{4} \forall \alpha$. Therefore in the Haldane phase the inaccessible entanglement is 2.  \\

To derive a similar result for all states in an MNC phase, we start from Eq.~\eqref{probs}. We first use symmetries of the left fixed point. The main step will be to use the simplified form of the projective representation in Eq.~\eqref{projrepmnc}. We begin by fixing the form of the left fixed point according to the symmetry imposed by Eq.~\eqref{projectiverep}. Inherited from the symmetries of the transfer matrix T, the left fixed point is also symmetric under $V(g) = \mathbb{1}\otimes\widetilde{V}(g)$, \textit{i.e.} 
$\rho_L =(\mathbb{1}\otimes\widetilde{V}(g))\rho_L(\mathbb{1}\otimes\widetilde{V}(g)^\dag)$.
By Schur's Lemma, $\rho_L$ can therefore be written as $\rho_L = \widetilde{\rho}_L \otimes \mathbb{1}$, for some matrix $\widetilde{\rho}_L$ carrying correlations and corresponding to the junk subspace.  We choose normalisation $\Tr(\rho_L) = 1$ which implies that $\Tr(\widetilde{\rho}_L) = 1/D_\omega$. \\
 
Inserting back the above form of the fixed point  $\rho_L$ into Eq.~\eqref{probs},  the irrep probabilities can be expressed as 
\begin{equation}
\begin{split}
      p_\alpha = \frac{1}{|G|} \sum_g & {\chi_\alpha(g)} \Tr( (\widetilde{\rho}_L \otimes  \mathbb{1}) (\mathbb{1} \otimes  {\widetilde{V}}(g))) \\
&\Tr( (\widetilde{\rho}_L\otimes  \mathbb{1}) (\mathbb{1} \otimes \widetilde{V}^\dag(g))). 
\end{split}
\end{equation}
Now multiplying out the corresponding parts of the tensor product, and reducing the trace over it,
\begin{equation}\label{probs1}
p_\alpha = \frac{1}{|G|} \sum_g  {\chi_\alpha(g)} \Tr( \widetilde{\rho}_L )^2 \Tr( \widetilde{V}(g))\Tr( \widetilde{V}^\dag(g)).
\end{equation}
Using now Eq.~\eqref{theorem} and that $\Tr(\widetilde{\rho}_L) = 1/D_\omega$, the above equation becomes
\begin{equation}  
p_\alpha = \frac{1}{|G|}\frac{1}{D_\omega^2} \sum_g \chi_\alpha(g) D_\omega^2 \delta_{g,e}. \end{equation} 

Using $\chi_\alpha(e) = 1$, $p_\alpha$ finally takes the same value for all irreps $\alpha$,
\begin{equation} p_\alpha = \frac{1}{|G|}. \end{equation} 
The irrep probabilities are degenerate, meaning that the weight of each sector contributes equally to $E_{inacc}$, and the entropy is maximised which confirms $E_{inacc}= \log(|G|)$. This demonstrates that MNC phases have the maximal set of topologically protected degenerate edge modes $D_\omega = |G|$ from Eq.~\eqref{edgemodes}, which is reflected in the inaccessible entanglement. 

	\subsection{Generalisation of bounds to non-MNC phases \label{nonmncsec}}

We generalise the inaccessible entanglement to describe all SPT phases, and crucially show that for non-MNC phases there exists a non-zero lower bound, 
\begin{equation} \label{bound}
\log(\frac{|G|}{|k|})  \leq E_{inacc} \leq \log(|G|),
\end{equation} for the group $G$ and projective centre group $k$ defined in Eq.\ \eqref{projcentre}.
We will first outline the steps taken to prove this. We use the general form of $V(g)$ in Eq.~\eqref{projgeneral}, and thus are prevented from easily evaluating the trace over $\rho_L$. However, much of the following proof remains in the same vein as in the previous Section, by simplifying the expression for $p_\alpha$ as much as possible by symmetry under $V(g)$. Finally we make an argument with entropy configurations from which we deduce the lower bound. \\

First we use the symmetry of $\rho_L$ under $V(g)$ to deduce the fixed point. By Schur's Lemma, the fixed point is again $\rho_L = \widetilde{\rho}_L \otimes \mathbb{1}$, where now  $\widetilde{\rho}_L =  \bigoplus_{a} \widetilde{\rho}_{L,a}$,  and each $\widetilde{\rho}_{L,a}$ has dimension $n_a \times n_a$. Again we choose the normalisation $\Tr(\widetilde{\rho}_L) = 1/D_\omega$ for convenience. Recalling the form of $V(g)$ from Eq.~\eqref{projgeneral}, the measurement probability then is written
\begin{equation} 
\begin{split}
p_\alpha = \frac{1}{|G|} \sum_g &\chi_\alpha(g) \Tr( \widetilde{\rho}_L \bigoplus_{a} \mu_a(g)\mathbb{1}_{n_a}  \otimes \widetilde{V}(g)) \\
&\Tr( \widetilde{\rho}_L \bigoplus_{a}\overline{\mu_a(g)}\mathbb{1}_{n_a}  \otimes \widetilde{V}^\dag(g)).
\end{split}
\end{equation} 
Separating the trace over the tensor product,
\begin{equation} 
\begin{split}
p_\alpha = \frac{1}{|G|} \sum_g & \chi_\alpha(g) \Tr( \widetilde{\rho}_L \bigoplus_{a} \mu_a(g)) \\
&\Tr( \widetilde{\rho}_L \bigoplus_{{a'}} \overline{\mu_{a'}(g)})\Tr( \widetilde{V}(g))\Tr( \widetilde{V}^\dag(g)).
\end{split}
\end{equation} 
Using Eq.~\eqref{theorem},  we can evaluate the traces over the projective representation, such that $p_\alpha$ further simplifies to
\begin{equation}
p_\alpha = \frac{D_\omega^2}{|G|}  \sum_{s \in k} \overline{\chi_\alpha(s)} \Tr( \widetilde{\rho}_L \bigoplus_{a} \mu_a(s))\Tr( \widetilde{\rho}_L \bigoplus_{a'} \overline{\mu_{a'}(s)}). 
\end{equation}
Crucially, at this point, $p_\alpha$ depends only on the value that $\chi_\alpha$ takes in $k\subset G$. This will lead to degeneracy in the different $p_\alpha$, giving our lower bound of $E_{inacc}$. To derive a simplified expression for $p_\alpha$, we insert $\widetilde{\rho}_L \bigoplus_{a} \mu_a(g) =  \bigoplus_{a}  \mu_a(g) \widetilde{\rho}_{L,a}$, and $D_\omega = \sqrt{|G|/|k|}$ to arrive at
\begin{equation}\label{eq:2}
p_\alpha = \frac{1}{|k|} \sum_{s \in k} \chi_\alpha(s) \sum_a \mu_a(s) \Tr(  \widetilde{\rho}_{L,a}) \sum_{a'}  \overline{\mu_{a'}(s)} \Tr( \widetilde{\rho}_{L,a'}). 
\end{equation}
We now use that the $\mu_a(g)$ are linear characters of $G$, \textit{i.e.}\ there is a function $u(a)$ such that $\mu_a(g)$ = $\chi_{u(a)}(g)$ $\forall \ g\in G$. Then, we can write $\chi_\alpha(s) \mu_a(s)\overline{\mu_{a'}(s)} = \chi_\alpha(s)\chi_{u(a)}(s)\overline{\chi_{u(a')}(s)}$. Combining the first two characters as $\chi_\alpha(g)\chi_{u(a)}(g) = \chi_{\alpha \cdot u(a)}(g)$, we can now use the row character orthogonality relation $\sum_{g \in G} \overline{\chi_\alpha(g)}\chi_\beta(g) = |G| \delta_{\alpha \beta}$ to simplify the sum $\sum_{s \in k}  \chi_{\alpha \cdot u(a)}(g) \overline{\mu_{a'}(s)} = |k| \delta_{\alpha\cdot u(a),u(a')} $\cite{karpilovsky1994group}. This leaves us with the final expression:
\begin{equation}\label{eq:3}
p_\alpha = \sum_a \Tr(\widetilde{\rho}_{L,a})  \Tr(\widetilde{\rho}_{L,u^{-1}(\alpha\cdot u(a))}). 
\end{equation}

We hereby arrive at the final expression above in Eq.~\eqref{eq:3}. To calculate the lower bound on inaccessible entanglement we use the fact that the probabilities $p_\alpha$ depend only on the values that $\chi_\alpha$ takes in $k$, which means that each value of $p_\alpha$ occurs $\frac{|G|}{|k|}$ times. When $p_\alpha$ have such a degeneracy, the configuration with lowest possible entropy is that for which $\frac{|G|}{|k|}$ of the $p_\alpha$ are equal to $\frac{|k|}{|G|}$, and the rest are 0. This leads to the lower bound on $E_{inacc}  = \log(\frac{|G|}{|k|})$ for non-MNC phases. Combining this with the upper bound given by the size of the symmetry group, leads to the main result presented in Eq.~\eqref{bound}.\\

\subsection{\label{stringorder}Comparison with string order parameter}

We show here that the string order parameter for 1D SPT phases \cite{pollmann2012detection} can be related to the inaccessible entanglement $E_{inacc}$. The probabilities $p_\alpha$ for each symmetry sector $\alpha$ are the Fourier transform of the string order parameter with identities as the end operators,
\begin{equation} \label{string}
    p_\alpha = \frac{1}{|G|} \sum_g \chi_\alpha(g) s(g,\mathbb{1},\mathbb{1}),
\end{equation} where the string order is defined
\begin{equation}
     s\left(g,O^A,O^B\right) = \Tr(\rho_L O^A V(g)) \Tr(\rho_L O^B V(g)^\dag), 
\end{equation} and $O^A,O^B$ are in a set of carefully chosen operators specific to the symmetry and projective representation which allow a particular selection rule to be detected. The rule changes at the phase transition point and allows unique determination of the phase, even distinguishing different MNC phases \cite{den1989preroughening,pollmann2012detection}. \\

This has allowed us to give an operational interpretation to string order as well as more concretely linking $E_{inacc}$ to phase detection. The expression in Eq.~\eqref{string} clarifies why inaccessible entanglement is not directly an order parameter; the missing end operators in inaccessible entanglement mean that the full information of the phase cannot be captured due to the lack of additional structure afforded, which the string order still possesses.

\section{\label{investigation}Investigation of the bounds}

In this Section we explore the bounds on the inaccessible entanglement in SPT states, and present numerical results to consolidate our findings. We consider the following main questions. Firstly, can states be constructed to explore the whole bound? Secondly, where do states typically live in the bound? We consider how to explicitly construct states both on the lower and upper bound, and probe the SPT phase and entanglement properties of the state. Lastly, we examine the effect of varying bond and physical dimension on the inaccessible entanglement. \\ 

We first introduce some conventions we will use in this Section. When performing interpolations between states of interest, we will filter certain components of the state $\ket{\Psi[A(\lambda)]}$, i.e.\  we choose $A(\lambda)$ in the form
\begin{equation}\label{eq:interpolation_tensor}
  A(\lambda) = \sum_{j\in I} A^{j} \ket{j} + \lambda \cdot\sum_{j\notin I} A^j \ket{j} ,  
\end{equation}
for some set of indices $I$. Here and in the rest of this Section, the basis set $\{\ket{j}\}$ is that which diagonalises $u(g)$, such that $u(g)\ket{j} = \chi_j(g)\ket{j}$. The path defined by tuning parameter $\lambda \in [0,1]$ for the state $\ket{\Psi[A(\lambda)]}$ is a particular adiabatic evolution where each MPS is the ground state of some Hamiltonian $H(\lambda)$. By interpolating $\lambda$ acting on a particular symmetry sector we can filter out the state's support on that sector, and effectively eliminate that symmetry action, which enables us to manipulate the symmetry of the state in a smooth, controlled manner. \\

Throughout this numerical investigation we mainly employ an  MPS construction which generates random states in a particular SPT phase by inputting $u(g)$ and $V(g)$ and symmetrising a random MPS tensor. This allows us to construct generic families of states in a chosen SPT phase and to study their properties. We first generate the random, injective tensor $M$ sampled from a Gaussian distribution of mean 0 and variance 1 over the complex numbers. We then sum over $M^i$ with elements of the symmetry as follows,
\begin{equation} \label{symconstruct}
    A^i = \sum_g \sum_j u(g)_{ij} \left( V(g) M^j V(g)^\dag \right).
\end{equation}
The MPS tensor defined above naturally satisfies the symmetries in Eq.~\eqref{fundtheorem} imposed by the fundamental theorem. The two crucial ingredients of the construction are the physical representation of the symmetry $u(g)$, written as a direct sum over its linear irreps $\chi_\alpha(g)$ with multiplicity $m_\alpha$, and the projective representation $V(g)$, built from projective irreps $V_a(g)$ with multiplicity $n_a$. The dimension of $V(g)$ is the bond dimension $D = \sum_a n_a \dim(V_a(g))$ while the dimension of $u(g)$ is the physical dimension $d = \sum_\alpha m_\alpha$. Note that all figures presented will display a sample of $10^6$ states.

\subsection{Tight bound for the inaccessible entanglement \label{tightbound}}

We demonstrate that there exists no tighter bound for $E_{inacc}$ in the trivial phase by interpolating from an example in an SPT-trivial phase to a product state, which adiabatically connects the upper and lower bound, as shown in Fig.~\ref{wholebound1}. We will use a toy MPS construction as our example for an SPT-trivial maximally entangled state with zero accessible entanglement. With this construction, we can interpolate to a product state and hence demonstrate the tightness of the bound. The MPS tensor is given by \begin{equation}\label{construction_triv}A^{(ij)}_{ij}= \ket{i} \bra{j},
\end{equation} 
where the indices $i$ and $j$ run from $1$ to $D$, and the physical index $(ij)$ runs from $1$ to $d=D^2$. This state is a product of maximally entangled pairs. The MPS generated by this tensor satisfies the symmetry condition of Eq.~\eqref{fundtheorem} with $u(g) = V(g) \bigotimes \overline{V(g)}$ and therefore $V(g)$ can be any invertible matrix satisfying the group relations of $G$. For convenience we will use $V(g)=\bigoplus_{a=0}^D \chi_a(g)$ to study the trivial SPT phase. Then, by the Clebsch-Gordan series the linear representation is written $u(g)=\bigoplus_{\alpha=0}^D \left( \mathbb{1}_D \otimes \chi_\alpha(g) \right)$.  \\

We consider $\mathbb{Z}_2 \times \mathbb{Z}_2$ symmetry by using the MPS tensor introduced in Eq.~\eqref{construction_triv} with bond dimension $D=4$ and physical dimension $d=16$. The path we choose is the interpolation $\ket{\Psi[A(\lambda)]}$ as defined in Eq.~\eqref{eq:interpolation_tensor} with  $I = \{0\}$ to filter all but the trivial irrep which connects the highly entangled state $\ket{\Psi[A(\lambda=1)]}$ to the product state $\ket{\Psi[A(\lambda=0)]}$ in the trivial irrep, while respecting the symmetry. This interpolation smoothly connects $E_{inacc}= 2$ to $E_{inacc} = 0$ tuning through an irrep probability distribution $p_\alpha = \{ \frac{1}{4},\frac{1}{4},\frac{1}{4},\frac{1}{4}\}$ to $p_\alpha=\{ 1,0,0,0\}$. The example state we give is but one of many possible examples which could have produced this. As we will later show, indeed states typically saturate the upper bound, while having (close to) zero inaccessible entanglement or residing in the middle region is rare. 

\subsection{Typical behaviour in the MNC and trivial phase \label{mncphase_typical}}

In this Section we discuss the inaccessible entanglement of states in the trivial and non-trivial (MNC) SPT phase of $\mathbb{Z}_2 \times \mathbb{Z}_2$. We begin by discussing the cluster state as a canonical example of a state in an MNC phase (the cluster phase). We comment on the allowed irrep probabilities in trivial and non-trivial SPT phases in this simple case; trivial order has no restrictions, whereas the MNC phase is fixed. \\

We first more exhaustively discuss the inaccessible entanglement of the cluster state, which we have calculated in Section \ref{maxinacc}. The cluster state has $p_\alpha = 1/4$ $ \forall \alpha$, so $E_{inacc} = 2$ which is the upper bound on inaccessible entanglement for this symmetry group. The cluster state is the unique ground state of the stabiliser Hamiltonian $H_C= - \sum_n \sigma^z_n \sigma^x_{n+1} \sigma^z_{n+2}$ for sites $n$, and it can be prepared by a finite depth circuit of $CZ =diag(1,1,1,-1)$ gates between neighbouring pairs of sites each initialised in the state $\ket{+}^{\otimes N}$; hence it is short-range entangled \cite{briegel2001persistent,raussendorf2003measurement,son2012topological}. An MPS description of the cluster state can be given by the Pauli matrices $\sigma^i$, $i= 1,x,y,z$ which has bond dimension $D=2$ and physical dimension $d=4$, so the linear representation has multiplicity $m_\alpha=1$ for each irrep such that $u(g) = \bigoplus_\alpha \chi_\alpha(g)$. Note that we use the notation $\sigma^i$ for MPS constructions and $\sigma(g)$ for representations, where both refer to the set of Pauli matrices. The Schmidt values of the state are $(\frac{1}{2},\frac{1}{2})$, so $E(\rho) = 2 S(\rho_A) = 2$. Therefore the accessible entanglement is zero; all the entanglement is inaccessible. Note that the factor of two comes from periodic boundary conditions; partitioning the system cuts through two bonds. \\
\begin{figure}[t]
\begin{center}
\includegraphics[scale=0.5]{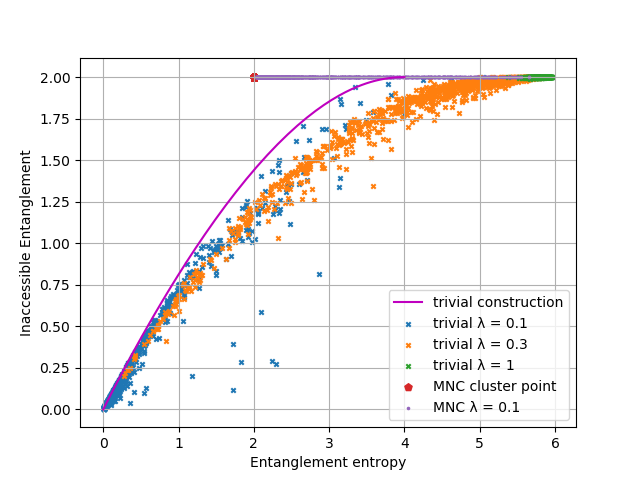}
\caption{\label{wholebound1} Inaccessible entanglement of states $\ket{\Psi[A(\lambda)]}$ for a filter $\lambda$. This is implemented in two ways: firstly, for families of random states and a certain $\lambda$ (scattered dots) defined by the MPS tensor in Eq.~\eqref{symconstruct}, and secondly an interpolation in $\lambda$ on a particular trivial MPS construction defined in Eq.~\eqref{construction_triv} (solid magenta line). The tensors are constructed with $\mathbb{Z}_2\times \mathbb{Z}_2$ symmetry and the interpolation on any MPS tensor $A(\lambda)$ as defined in Eq.~\eqref{eq:interpolation_tensor}, where $I = \{0\}$ is the trivial irrep. For reference, the cluster state is also displayed (red dot).}
\end{center}
\end{figure}

Let us now consider the rest of the cluster phase. The inaccessible entanglement persists throughout the cluster phase, as demonstrated in Fig.~\ref{wholebound1} by studying random states with a non-trivial multiplicity in the physical representation. In this figure we use the MPS tensor $A$ from Eq.~\eqref{symconstruct}, where SPT-trivial states are constructed with $n_a = 4, \hspace{1ex}m_\alpha =2 \hspace{1ex}\forall\hspace{1ex} a,\alpha$ (bond dimension $D=8$ and physical dimension $d = 16$), while the MNC phase is constructed with $n_a = 4, \hspace{1ex}m_\alpha =2 \hspace{1ex}\forall\hspace{1ex} a,\alpha$ ($D=4$ and $d = 16$). We can view $E_{inacc}$ as an invariant in a particular sense since it takes the same value throughout this phase, being independent of the correlations within the junk subspace and only dependent on the part of the state in which symmetries act non-trivially. States in the cluster phase with larger bond dimension can allow more entanglement due to a non-trivial junk subspace, which allows non-zero accessible entanglement. These results provide a new angle on the cluster phase. \\

Let us compare the trivial phase to the MNC phase. Fig.~\ref{wholebound1} demonstrates that trivial SPT order displays less restricted combinations of $(E_{inacc},E)$ compared to the MNC phase; this originates from a lack of enforced structure in irrep probabilities, while the MNC phase has a fixed, degenerate probability distribution. Despite this key difference, a random state in the SPT trivial phase typically has similar $(E_{inacc},E)$-values compared to the MNC phase value, as they are generally close to the upper bound. The closer we tune random states in either phase towards the product state, the lower the entanglement gets on average, but while for the trivial phase the inaccessible entanglement also decreases on average, the MNC phase has a fixed inaccessible entanglement. Since the entanglement entropy can't decrease below the fixed $E_{inacc}$ in the MNC phase, the path tuning an MNC phase to a product state would see a phase transition. \\

\begin{figure}[t]
\begin{center}
\includegraphics[scale=0.5]{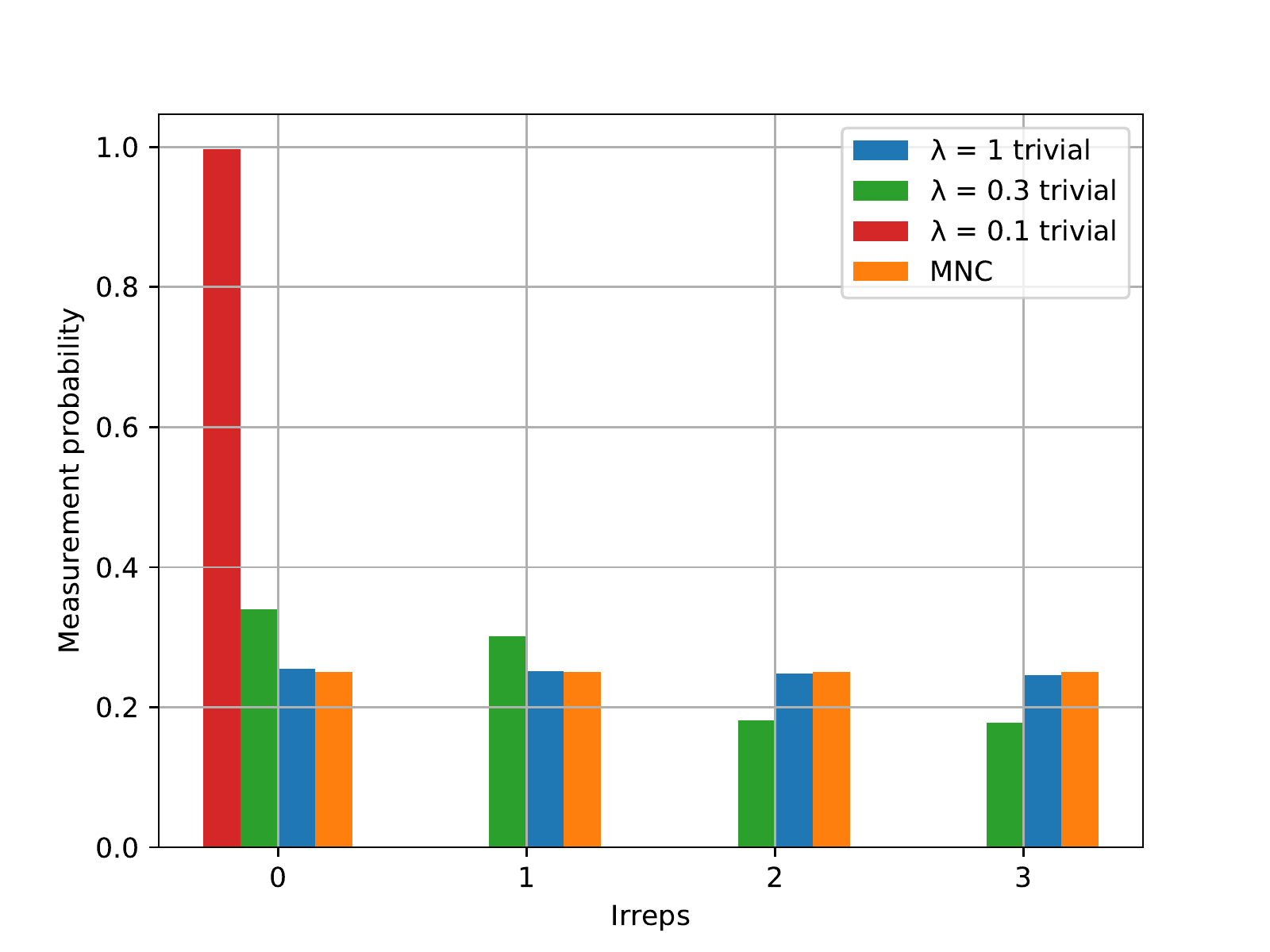}
\caption{\label{probsz2z2} Irrep measurement probability distributions for a selection of four random states from the families displayed in Fig.~\ref{wholebound1}, three with trivial SPT order with a filtering $\lambda$, where $\lambda \rightarrow 0$ denotes tuning the state towards a product state, and one with MNC order. The MNC phase strictly contains states with a degenerate irrep distribution, whereas trivial states can have entirely different irrep probabilities. The interpolation is from an almost degenerate irrep distribution, which mimics the MNC phase, to an irrep distribution weighted only on the trivial irrep, corresponding to zero inaccessible entanglement; notice that between the $\lambda = 1$ state and the MNC phase the probabilities are not quite equal due to the different SPT order.}
\end{center}
\end{figure}

We now explore the irrep probability distributions, which can reveal crucial information about the composition of the inaccessible entanglement, as displayed in Fig.~\ref{probsz2z2}. While the MNC phase can only be realised by a degenerate probability distribution, the trivial phase is unconstrained. Although typically random trivial states have near maximal inaccessible entanglement, one can distinguish it from an MNC ordered state given enough precision on the irrep probabilities. Additionally, we emphasise that while the MNC phases of a system with a symmetry $G$ have a fixed value for the inaccessible entanglement, this is not enough to distinguish two MNC phases from the same group $G$. While two such phases are mathematically inequivalent, some of their physical properties, such as the topologically protected edge mode degeneracy and entanglement spectrum, are unchanged.
\subsection{Exploring a non-MNC phase}

How does the inaccessible entanglement behave for non-MNC phases? In this part of our Investigation, we explore the properties of non-MNC phases, which have interesting characteristics compared to MNC or trivial phases, since states with this SPT order live in a finite region of the bound which is fully restricted, having a non-zero, symmetry dependent lower and upper bound, which are not equal to each other. We study the simplest symmetry hosting a non-MNC phase $G = \mathbb{Z}_4 \times \mathbb{Z}_2$, which has $H^2(G,U(1)) = \mathbb{Z}_2$. \\

The first point of interest is the possibility to drive states towards the lower bound by reducing the symmetry of the state to effective sub-symmetries. Secondly, we observe very little difference in the inaccessible entanglement between the trivial and non-trivial phase for random states; they both reside near the upper bound. As before, the trivial and non-trivial phases are still discriminated by the structure of irrep measurement probabilities. Finally, we make an analysis on the effect of bond dimension on the patterns of inaccessible entanglement in random states, and show that increasing $D$ increases the variance and mean of the inaccessible entanglement. We will clarify that this is due to a significant dependence on the structure of the junk subspace. 

\begin{figure}[t]
\begin{center}
\includegraphics[scale=0.6]{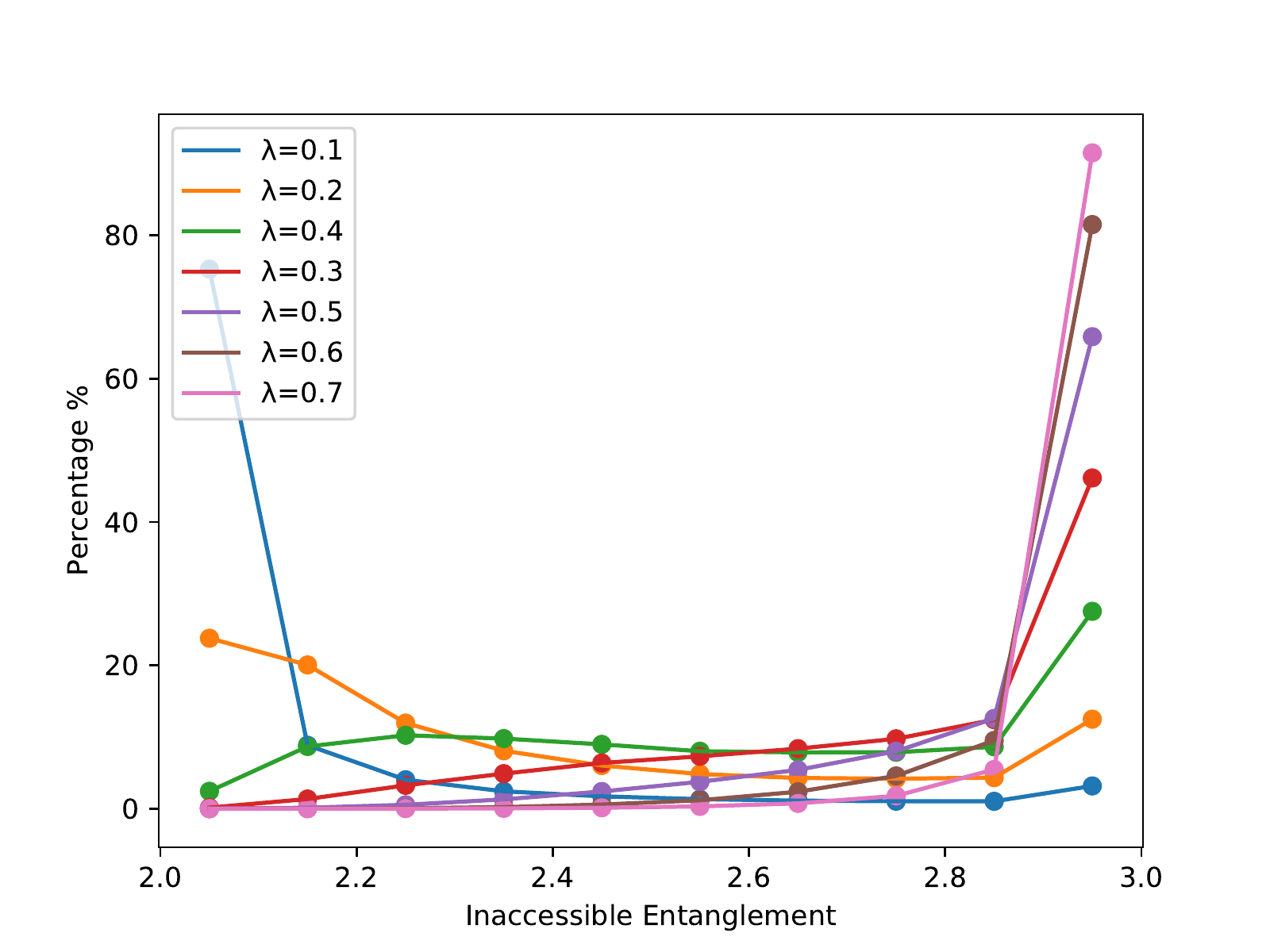}
\caption{\label{lowerbound} Histogram of the inaccessible entanglement for the non-MNC phase of $\mathbb{Z}_4 \times \mathbb{Z}_2$ in families of the random MPS tensor A with a filter $\lambda$ which drives states towards an effective $\mathbb{Z}_2 \times \mathbb{Z}_2$ symmetry as $\lambda \rightarrow 0$. Moving through each $\lambda$ value illustrates how states are gradually forced from the upper bound onto the lower bound, which corresponds to the upper bound of the trivial phase with $\mathbb{Z}_2 \times \mathbb{Z}_2$ symmetry.}
\end{center}
\end{figure}
\subsubsection{Reduced effective symmetries}

While in Section \ref{mncphase_typical} we showed that states in the trivial phase can be driven to the lower bound ($E_{inacc} = 0$ for trivial states) simply by interpolating towards a product state, we now show that considering a larger sub-symmetry allows additional interesting behaviour. Indeed, both the trivial and the non-MNC phase can be driven towards $E_{inacc} = 2$ by effectively enforcing a $\mathbb{Z}_2 \times \mathbb{Z}_2$ symmetry through suppression of particular $A^i$ in the MPS tensor $A$ as introduced in Eq.~\eqref{construction_triv}, with $u(g) = \bigoplus_\alpha \chi_\alpha(g)$ where $m_\alpha=1$ $\forall \alpha$, and with $V(g)$ constructed with $n_a = 2 $ for $a \in \{0,1\}$ since there are two projective irreps. We refer to $E_{inacc} = 2$ as the lower bound, although it is only the lower bound for the non-trivial phase. \\

The non-MNC phase is depicted in Fig.~\ref{lowerbound}. As $\lambda \rightarrow 1$, random states typically live near the upper bound $E_{inacc} = 3$ in either phase, while as $\lambda \rightarrow 0$ states get trapped nearer the lower bound. The latter happens at comparatively low values of $\lambda$, illustrating further that typical states tend to have high inaccessible entanglement. As an aside, we do not include the value $\lambda = 0$ for which the symmetry is actually enforced,  in order to be consistent with assumptions made in the derivation of the bounds, since these states will not be injective.\\

Let us describe how we determine which irreps are filtered. We denote elements $g$ of $\mathbb{Z}_4\times\mathbb{Z}_2$ by $g=(x,y)$ where $x=0,1,2,3$ and $y=0,1$. Then, the irreps can be labelled by a pair $\alpha=0,1,2,3$ and $\beta=0,1$ such that 
\begin{equation}
    \chi_{\alpha,\beta}(x,y) = (i)^{\alpha x}(-1)^{\beta y}.
\end{equation}

The irreps we filter are those for which $\alpha = 1$ or 3, such that the unfiltered irreps are only faithful to a $\mathbb{Z}_2\times\mathbb{Z}_2$ subgroup generated by $(x,y) = (2,0)$ and $(0,1)$. Therefore in the limit $\lambda \rightarrow 0$ the MPS effectively only has a $\mathbb{Z}_2\times\mathbb{Z}_2$ symmetry.

 \subsubsection{Comparing trivial to non-trivial phase \label{sat}}
 
We discuss how the inaccessible entanglement is characterised in the non-MNC and trivial phase of $\mathbb{Z}_4 \times \mathbb{Z}_2$, as illustrated by the following numerical results. We show that SPT trivial states generally have high inaccessible entanglement, which makes them hard to distinguish from the non-trivial states, but the lack of degeneracy in the irrep probabilities  (present in the non-trivial phase) reveals their triviality. The SPT-trivial states are constructed with $n_a = 1, \hspace{1ex}m_\alpha =2 \hspace{1ex}\forall\hspace{1ex} a,\alpha$ (bond dimension $D=8$ and physical dimension $d = 16$), while the non-MNC phase is constructed with $n_a = 2, \hspace{1ex}m_\alpha =1 \hspace{1ex}\forall\hspace{1ex} a,\alpha$ ($D=8$ and $d = 8$).\\

\begin{figure}[t]
\begin{center}
\includegraphics[scale=0.6]{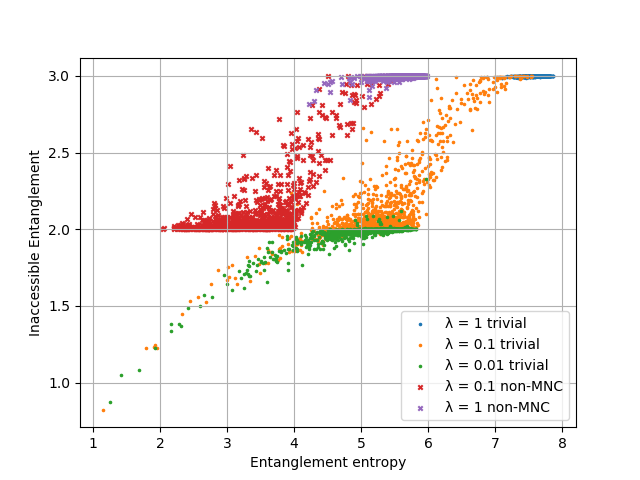}
\caption{\label{wholeboundz4z2} Inaccessible entanglement versus entanglement of states in the SPT phases of $G = \mathbb{Z}_4 \times \mathbb{Z}_2$.  We study random states $\ket{\Psi[A(\lambda)]}$ generated with the construction in Eq.~\eqref{symconstruct} with a particular filtering $\lambda$ effecting the sub-symmetry $\mathbb{Z}_2 \times \mathbb{Z}_2$ as in Fig.~\ref{lowerbound}, acting on three trivial and two non-MNC families of states. The non-MNC phase is restricted to within the bound $2\leq E_{inacc} \leq 3$ whereas the trivial phase is lower bounded by zero.}
\end{center}
\end{figure}

Studying Fig.~\ref{wholeboundz4z2}, there are some main remarks. From Eq.~\eqref{bound}, the non-trivial SPT phase is bounded by $2 \leq E_{inacc} \leq 3$ whereas the trivial SPT can get arbitrarily close to zero inaccessible entanglement. Since states in the trivial phase can reach values below $E_{inacc}=2$, decreasing $\lambda$ drives states towards approaching this value from below, rather than from above. States with non-MNC order meanwhile become increasingly clustered above the bound, as they can't pass below it, as we saw earlier displayed in Fig.~\ref{lowerbound}.\\

\begin{figure}[t]
\begin{center}
\includegraphics[scale=0.6]{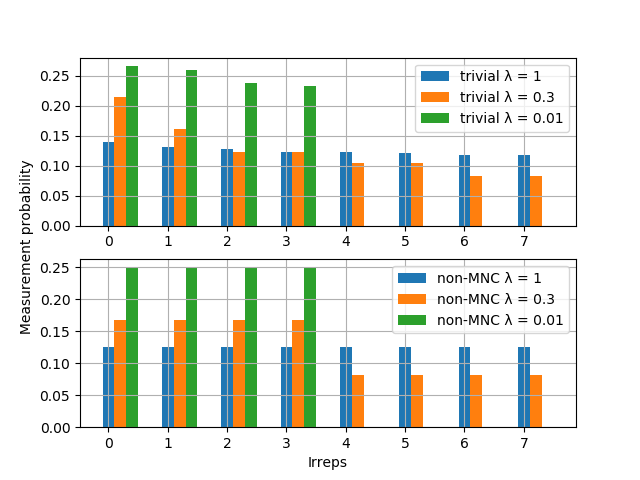}
\caption{\label{probsz4z2} Irrep measurement probabilities of a selection of random states picked from distributions as in Fig.~\ref{wholeboundz4z2} with filtering $\lambda$ where $\lambda \rightarrow 0$ drives states towards an effective $\mathbb{Z}_2 \times \mathbb{Z}_2$ symmetry, comparing the trivial phase to the non-MNC phase of $\mathbb{Z}_4 \times \mathbb{Z}_2$. The non-MNC phase irrep probabilities are four-fold degenerate for any filtering, whereas the trivial phase gives different probabilities to each irrep in general. However, for both phases, as $\lambda \rightarrow 0$ the irrep probabilities increasingly resemble the degeneracy of the MNC phase of $\mathbb{Z}_2 \times \mathbb{Z}_2$.}
\end{center}
\end{figure}

We demonstrate that one can actually infer SPT triviality or non-triviality from the irrep probability distributions, displayed in Fig.~\ref{probsz4z2}. Notice that non-MNC phase irrep probabilities are four-fold degenerate; this is due to only two inequivalent irreps which contribute to the calculation for the measurement probabilities $p_\alpha$. By contrast each irrep probability can be different in the trivial phase. \\

Let us consider the physical insights offered by three example parameters. In each, the difference between trivial and non-trivial phase is apparent. Firstly, recall from Section \ref{sat} that all states with $\lambda = 1$ have a near maximal inaccessible entanglement. The closer to saturating this value a state is, the closer to degenerate the irrep probabilities become. Therefore in the non-MNC phase, the probabilities are approximately $p_\alpha = \frac{1}{8} \forall \alpha$. Notably, this still differs from the behaviour in an MNC phase, as the irrep probabilities are still not completely flat, having instead a four-fold degeneracy. The degeneracy in the non-MNC phase becomes more apparent for $\lambda = 0.3$, and even more prominently with $\lambda = 0.01$ which generates the effective sub-symmetry, and echoes the irrep probabilities of the cluster phase, where $p_\alpha = \frac{1}{4}$ for half the irreps $\alpha$, with the other half being zero. For all three examples, the trivial phase may have eight different irrep probabilities, which are quite similar to the non-trivial probabilities but are very unlikely to be exactly degenerate; although such states may have indistinguishable inaccessible entanglement on average compared to the non-trivial phase, it is remarkable that due to a trivial projective representation characterising the state, they lose crucial symmetry structure in the irrep probabilities. 
\subsubsection{Accessible entanglement in the junk subspace}

\begin{figure}[t]
\begin{center}
\includegraphics[scale=0.6]{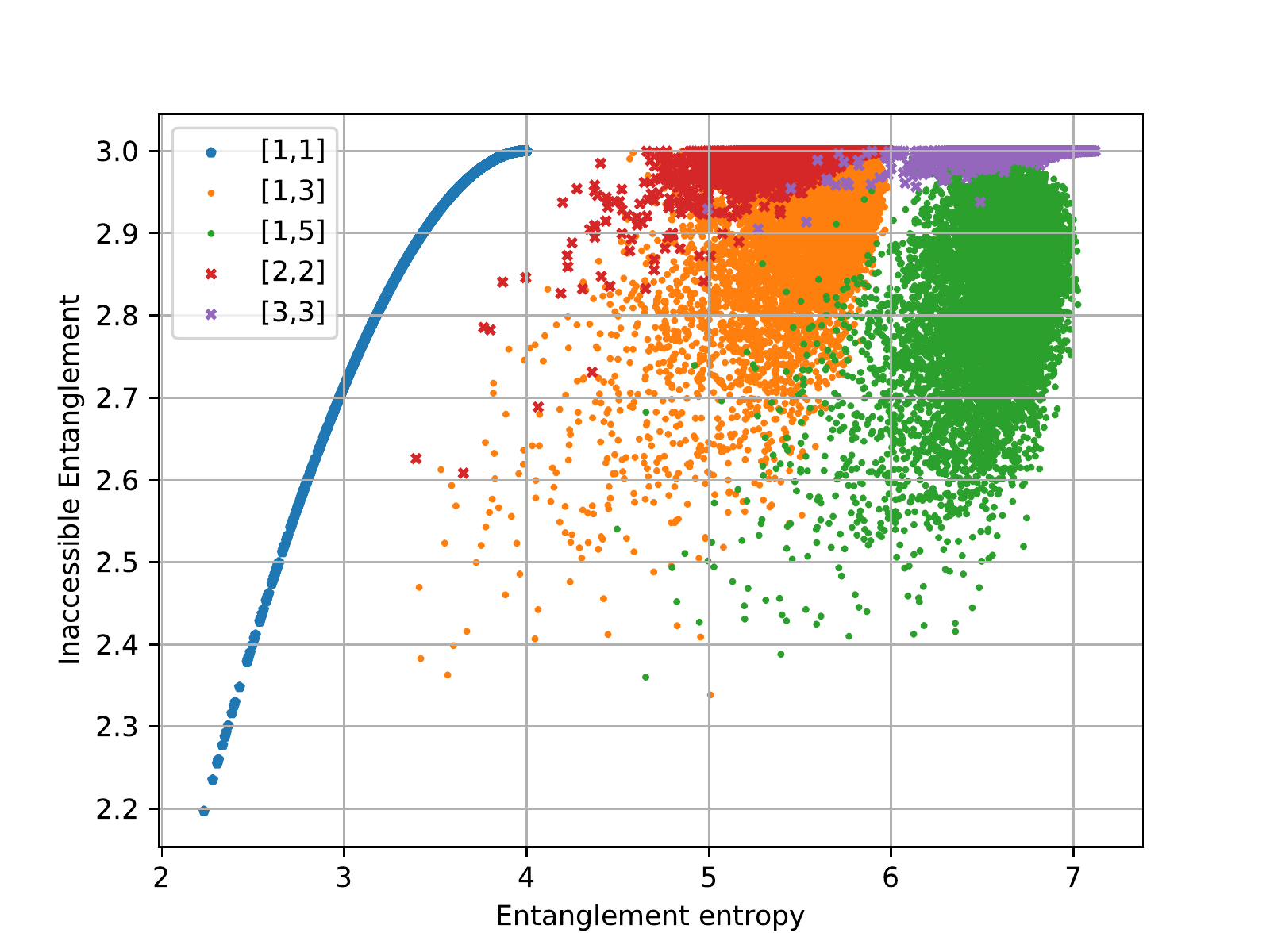}
\caption{\label{bonddimcheck} Effects of different irrep distributions of $V(g)$ for random states constructed as in Eq.~\eqref{symconstruct} in the non-MNC phase of $\mathbb{Z}_4 \times \mathbb{Z}_2$, with a constant $u(g) = \bigoplus_\alpha \chi_\alpha(g)$ where $m_\alpha=1$ $\forall \alpha$. The labels $[n_0,n_1]$ are the multiplicities of the two irreducible projective representations in $V(g)$. The first example (blue dots) has a trivial junk subspace and is therefore limited to a line. States with uneven multiplicities on the irreps have less possible inaccessible entanglement compared to the same bond dimension which has equal multiplicities. Larger bond dimension allows more entanglement but the inaccessible entanglement remains capped at $E_{inacc} = 3$, as this upper bound depends only on $G$. }
\end{center}
\end{figure}

We examine how inaccessible entanglement depends on the bond dimension. Clearly increasing bond dimension shifts states to a larger entanglement entropy by virtue of the tensor network construction, but it also leads to a higher average inaccessible entanglement. We consider the non-trivial SPT phase of $\mathbb{Z}_4 \times \mathbb{Z}_2$, which has two projective irreps. We argue that an imbalance in the multiplicities of the two irreps leads to larger proportions of states near the upper bound of inaccessible entanglement compared to equal multiplicities. The numerics for this study are displayed in Fig.~\ref{bonddimcheck}, where we construct states as in Eq.~\eqref{symconstruct} with a constant $u(g) = \bigoplus_\alpha \chi_\alpha(g)$ where $m_\alpha=1$ $\forall \alpha$ but where we vary the multiplicities of $V(g)$ and hence the junk subspace. \\

Let us first explain the significance of the junk subspace, introduced in Eq.~\eqref{projrepmnc}, by considering the simple case of the MNC phase. The cluster state is represented by the MPS tensor $A^i = \sigma^i$ which has a trivial junk subspace. In the MNC phase, the accessible entanglement is determined only by the junk subspace of the MPS tensors; a trivial junk subspace has $E_{acc} = 0$, and non-trivial subspace allows non-zero accessible entanglement. A large junk subspace, effected by higher bond dimension, has the potential to host more entanglement since the upper bound on entanglement entropy is given by the Schmidt rank which is $\log(D)$; in other words, there are more free parameters and hence more correlations are possible. By increasing the junk subspace and exploring the cluster phase $A^i = B^i \otimes \sigma^i$ for some non-trivial $B^i$, the inaccessible entanglement remains the same, since it depends only on the part of the state which transforms non-trivially under the symmetry, yet the entanglement and accessible entanglement can grow. We now extend this argument to a more general case. \\

In non-MNC phases, as opposed to MNC phases, the inaccessible entanglement is not fixed, so there is greater freedom in how the total entanglement is divided into accessible and inaccessible parts. However, the importance of the junk subspace for the accessible entanglement is still clear. As the total size of the junk subspaces in each block increases, so does the the possible values of the inaccessible entanglement for a given total entanglement. Indeed, the family $n=[1,1]$ has no junk subspace, and we see that the inaccessible entanglement is uniquely determined by the total entanglement. Comparing the two families $n=[1,5]$ and $n=[3,3]$ reveals the same effect. Even though total bond dimension is the same, the structure of the junk subspaces is different, with the number of free parameters in the junk subspace being $1 + 5 \times 5 = 26$ compared to $2 \times 3 \times 3 = 18$ respectively. In the former case, the larger number of parameters means more possible values of accessible entanglement, leading to the more spread-out distribution in Fig.~\ref{bonddimcheck}. This highlights the power of the inaccessible entanglement to also capture properties due to symmetry structure on the virtual level.

\section{Discussion \label{discuss}}

\subsection{Restricting symmetry to a subgroup}

Up until this point, we have fixed a symmetry group $G$ and considered the inaccessible entanglement under $G$-LOCC for different $G$-symmetric states. Now, we consider the inverted scenario, in which we fix a particular state and consider $G$-LOCC for different symmetries $G$. Intuitively, reducing the number of enforced symmetries should reduce $E_{inacc}$. To see this explicitly, let us take a state in an MNC phase of $G$ and determine $E_{inacc}$ under $H$-LOCC where $H$ is a subgroup of $G$. Following the calculation in Section \ref{maxinacc}, we find that $p_\alpha = \frac{1}{|H|}$ for all $\alpha$, where $\alpha$ now runs over the $|H|$ irreps of $H$. Therefore, Eq.~\eqref{inacc} gives $E_{inacc}=\log(|H|)$. As expected, we find that $E_{inacc}$ decreases as smaller symmetry groups are enforced, corresponding to more of the entanglement in our fixed state becoming accessible as our LOCC becomes less restricted.\\

Restricting symmetries to a subgroup also effects the SPT order of a state; a state with SPT order under $G$ symmetry may be trivial under $H$ symmetry for $H\subset G$. Yet, as demonstrated above, if we begin with a state in an MNC phase, the inaccessible entanglement under $H$-LOCC will always take the maximum value. For example, consider a state in an MNC phase of $G = \mathbb{Z}_4 \times \mathbb{Z}_4$. If we restrict to the subgroup $H = \mathbb{Z}_2 \times \mathbb{Z}_2$, the SPT order becomes trivial, but we nonetheless have $E_{inacc} = \log(|H|) = 2$. This provides an example of a state in a trivial SPT phase which, due to the ``hidden'' presence of SPT order under a larger symmetry group, maximises $E_{inacc}$ in the way that would normally be expected from a state in the MNC phase. The maximal inaccessible entanglement therefore captures the underlying MNC behaviour, even though the actual SPT order is destroyed by operations which do not protect the symmetry.

\subsection{Subsystem SPT order}

Our results immediately apply to certain 2D subsystem SPT (SSPT) phases \cite{you2018subsystem,stephen2019subsystem}. SSPT order generalises the concept of SPT order to include subsystem symmetries, which act on rigid subsystems such as lines or fractals. As an example, the 2D cluster state \cite{raussendorf2003measurement} has subsystem symmetries corresponding to flipping every spin along a line \cite{else2012symmetry,raussendorf2019computationally,you2018subsystem}. These systems can often be analysed with dimensional reduction which translates the subsystem symmetry group in 2D to an extensive global on-site symmetry group of an effective 1D system \cite{raussendorf2019computationally,devakul2018universal,stephen2019subsystem,Daniel2020computational}. Because of this, we can immediately apply our results to these phases as well.\\

For example, if we place the 2D cluster state on a cylinder of circumference $N$ such that the line symmetries wrap diagonally along the cylinder periodically, we can consider each block of $N\times N$ spins as a single site, such that we get an effective 1D chain with symmetry group $G=(\mathbb{Z}_2\times \mathbb{Z}_2)^N$ \cite{raussendorf2019computationally}. For each $N$, this 1D system is in an MNC phase with respect to $G$, so our general results can be applied to show that the inaccessible entanglement will be equal to $N$. For the 2D cluster state, this is all of the entanglement, which shows that there is no accessible entanglement in the state if the subsystem symmetries are enforced. The above analysis, which extends to the general family of fractal subsystem SPT phases defined in Refs. \cite{stephen2019subsystem,Daniel2020computational}, shows that $E_{inacc}$ for these subsystem SPT phases satisfies an area law in 2D.

\subsection{Relation to computational power}

The study of SPT order shares a remarkably symbiotic relationship with quantum computation via the paradigm of measurement-based quantum computation (MBQC) \cite{raussendorf2003measurement}. In MBQC, quantum computation is performed using single-site projective measurements on an entangled many-body state. A state is called a universal resource for MBQC if these measurements allow the circuit model of quantum computation to be efficiently simulated, with the cluster state being a canonical example. Characterizing universal resource states is a fundamental open problem in MBQC. Remarkably, there is a deep connection between the MBQC universality of a state and its SPT order: Many SPT phases in 1D and 2D are computational phases of matter, meaning that every state within the phase is a universal resource \cite{miyake2010quantum,prevWork1,miller2015resource,stephen2017computational,raussendorf2019computationally,devakul2018universal,stephen2019subsystem,Daniel2020computational}. A question which arises naturally due to their separate connections to SPT order is the following: can inaccessible entanglement predict whether a state is computationally universal for MBQC? \\

One intriguing connection comes from the importance of MNC phases in each setting: these are the phases which are universal for MBQC in 1D \cite{miyake2010quantum,prevWork1,stephen2017computational}, and they are also the phases in which $E_{inacc}$ is maximised in the entire phase. This connection extends to the 2D SSPT phases discussed in the previous Section \cite{raussendorf2019computationally,devakul2018universal,stephen2019subsystem,Daniel2020computational}. This suggests that a maximised value of $E_{inacc}$ may imply that a state is universal. However if we more closely analyze the MBQC scheme in Ref.~\cite{stephen2017computational}, which utilises 1D MNC SPT phases, we see that the connection might not always be certain. In each phase, there are certain measure-zero subsets of states which are not universal, even though $E_{inacc}$ remains maximised within these subsets. Of course, this does not preclude the existence of a different MBQC scheme for which these subsets are universal. Indeed, a similar scheme from Ref.~\cite{miller2015resource} comes with a different set of non-universal states which covers some of the holes left by the scheme of Ref.~\cite{stephen2017computational}. \\

When we consider non-MNC phases, the connection becomes less clear still. In Ref.~\cite{Stephen_2017}, a scheme of MBQC is introduced which utilises non-MNC phases. Crucially, unlike the MNC case, universality is not guaranteed in non-MNC phases, and is determined principally by the on-site symmetry representation $u(g)$. To check whether the universal states within a phase can be distinguished from the non-universal ones, we took random states from each case and computed their $E_{inacc}$. However, we found that there was no clear differentiating behaviour of $E_{inacc}$ between the two cases. This suggests that $E_{inacc}$ does not relate strongly to computational power beyond the MNC case. 

\section{\label{conclusion}Conclusion and Outlook}

We demonstrated that investigating the properties of $G$-LOCC is deeply connected to questions about phases of matter. We defined the accessible entanglement $E_{acc} = \sum_\alpha p_\alpha E_\alpha$ in this setting operationally, and thereby showed that the inaccessible entanglement corresponds to the number of Bell pairs inextricable under symmetry-restricted local operations. The main result of this work is that, given a 1D system with a global onsite (finite Abelian) symmetry, there is a tight bound on the inaccessible entanglement $\log(\frac{|G|}{|k|})  \leq E_{inacc} \leq \log(|G|)$, which depends on the symmetry group and SPT phase (via $k$) in question. This shows that there is always some entanglement present in non-trivial SPT phases which cannot be extracted via symmetry-respecting operations. In the maximally non-commutative SPT phases, which play an important role in measurement-based quantum computation \cite{prevWork1,stephen2017computational}, we have $|k|=1$, such that every state in these phases has $E_{inacc} =\log(|G|)$. In particular, the 1D and 2D cluster states, which are highly entangled by certain measures \cite{briegel2001persistent,verstraete2004entanglement}, have zero accessible entanglement under $G$-LOCC for suitable $G$. \\

We studied these bounds numerically by constructing random states in different SPT phases and calculating $E_{inacc}$. We find that the whole bound can be explored, although typical states tend to have near maximal $E_{inacc}$, even in the trivial phase. We characterised those states near the lower bound as those which have low weight in certain symmetry sectors, and therefore have an effective subgroup symmetry. Although it is difficult to distinguish trivial from non-trivial SPT phases by $E_{inacc}$ alone, we show that they can be distinguished by their irrep probabilities, which exhibit an exact degeneracy in non-trivial phases. We demonstrate that the inaccessible entanglement is the entropy of the irrep probability distribution corresponding to the Fourier transform of the string order parameter, which allows $E_{inacc}$ to harness some of the properties of the order parameter, and the irrep probabilities themselves are a good contender to access this information. Recent studies have accessed the entanglement spectrum of the cluster state on a noisy, intermediate-scale quantum (NISQ) computer \cite{choo2018measurement,2002.04620}, and we believe a similar study could be done with our method which might prove more able to identify topological phases beyond the cluster state and illuminate their computational power. \\

We also affirmed the question: can \enquote{latent} inaccessible entanglement appear due to the presence of a larger symmetry, which the description misses out on? We show that an MNC phase protected by some symmetry $G$ can become SPT-trivial by partially violating the symmetry through restricting operations to a symmetry $H \subset G$. This ``hidden'' order is detected by the inaccessible entanglement, since $E_{inacc}$ still contains the information of the SPT phase under the original symmetry that is not detected by the entanglement alone. \\

Throughout this work, we have focused on 1D SPT phases, as well as 2D subsystem SPT phases via dimensional reduction. For 2D SPT phases with global symmetries, the classification is given by the third cohomology group $H^3(G,U(1))$, and it is not clear what bounds may exist for the inaccessible entanglement. Therefore a natural future direction for investigation is to extend $E_{inacc}$ to PEPS.\\

An intriguing question is whether the inaccessible entanglement acts as a resource for a task in quantum information processing, in analogy to the super-selection induced variance (SIV) which can be viewed as a resource for quantum data hiding, introduced in Ref.~\cite{PhysRevA.70.042310}. It was found that states obeying particle number conservation can be distilled into a part containing pure SIV and a disjoint part containing pure entanglement, which we can now confirm is the accessible entanglement. The presence of SIV is a necessary condition for perfect data hiding, a protocol where classical data is hidden in such a way that it cannot be recovered by LOCC without quantum communication \cite{divincenzo2002quantum,verstraete2003quantum}. One can ask whether these non-local quantities can be formalised into the framework of a resource theory. It is also natural to ask about the connection between the inaccessible entanglement and SIV, by extending the definition of SIV to arbitrary symmetries. \\

We would like to thank Nick Bultinck, David Sauerwein, and Frank Verstraete for insightful discussions and comments. 
DTS was supported by a fellowship from the Natural Sciences  and  Engineering  Research  Council  of  Canada (NSERC). 
This work has been supported by the European Research Council (ERC) under the European Union’s Horizon 2020 research and innovation programme through the ERC-StG \mbox{WASCOSYS} (No.~636201) and the ERC-CoG GAPS (No.~648913), 
by the Deutsche Forschungsgemeinschaft (DFG) under Germany’s Excellence Strategy (EXC-2111 -- 390814868),
and by the Severo Ochoa project SEV-2015-0554 (MINECO). 

\bibliography{mylib.bib}	

\begin{thebibliography}{70}%
\makeatletter
\providecommand \@ifxundefined [1]{%
 \@ifx{#1\undefined}
}%
\providecommand \@ifnum [1]{%
 \ifnum #1\expandafter \@firstoftwo
 \else \expandafter \@secondoftwo
 \fi
}%
\providecommand \@ifx [1]{%
 \ifx #1\expandafter \@firstoftwo
 \else \expandafter \@secondoftwo
 \fi
}%
\providecommand \natexlab [1]{#1}%
\providecommand \enquote  [1]{``#1''}%
\providecommand \bibnamefont  [1]{#1}%
\providecommand \bibfnamefont [1]{#1}%
\providecommand \citenamefont [1]{#1}%
\providecommand \href@noop [0]{\@secondoftwo}%
\providecommand \href [0]{\begingroup \@sanitize@url \@href}%
\providecommand \@href[1]{\@@startlink{#1}\@@href}%
\providecommand \@@href[1]{\endgroup#1\@@endlink}%
\providecommand \@sanitize@url [0]{\catcode `\\12\catcode `\$12\catcode
  `\&12\catcode `\#12\catcode `\^12\catcode `\_12\catcode `\%12\relax}%
\providecommand \@@startlink[1]{}%
\providecommand \@@endlink[0]{}%
\providecommand \url  [0]{\begingroup\@sanitize@url \@url }%
\providecommand \@url [1]{\endgroup\@href {#1}{\urlprefix }}%
\providecommand \urlprefix  [0]{URL }%
\providecommand \Eprint [0]{\href }%
\providecommand \doibase [0]{http://dx.doi.org/}%
\providecommand \selectlanguage [0]{\@gobble}%
\providecommand \bibinfo  [0]{\@secondoftwo}%
\providecommand \bibfield  [0]{\@secondoftwo}%
\providecommand \translation [1]{[#1]}%
\providecommand \BibitemOpen [0]{}%
\providecommand \bibitemStop [0]{}%
\providecommand \bibitemNoStop [0]{.\EOS\space}%
\providecommand \EOS [0]{\spacefactor3000\relax}%
\providecommand \BibitemShut  [1]{\csname bibitem#1\endcsname}%
\let\auto@bib@innerbib\@empty
\bibitem [{\citenamefont {Bennett}\ \emph {et~al.}(1993)\citenamefont
  {Bennett}, \citenamefont {Brassard}, \citenamefont {Cr\'epeau}, \citenamefont
  {Jozsa}, \citenamefont {Peres},\ and\ \citenamefont
  {Wootters}}]{bennett1993teleporting}%
  \BibitemOpen
  \bibfield  {author} {\bibinfo {author} {\bibfnamefont {C.~H.}\ \bibnamefont
  {Bennett}}, \bibinfo {author} {\bibfnamefont {G.}~\bibnamefont {Brassard}},
  \bibinfo {author} {\bibfnamefont {C.}~\bibnamefont {Cr\'epeau}}, \bibinfo
  {author} {\bibfnamefont {R.}~\bibnamefont {Jozsa}}, \bibinfo {author}
  {\bibfnamefont {A.}~\bibnamefont {Peres}}, \ and\ \bibinfo {author}
  {\bibfnamefont {W.~K.}\ \bibnamefont {Wootters}},\ }\href {\doibase
  10.1103/PhysRevLett.70.1895} {\bibfield  {journal} {\bibinfo  {journal}
  {Phys. Rev. Lett.}\ }\textbf {\bibinfo {volume} {70}},\ \bibinfo {pages}
  {1895} (\bibinfo {year} {1993})}\BibitemShut {NoStop}%
\bibitem [{\citenamefont {Ekert}(1991)}]{ekert1991quantum}%
  \BibitemOpen
  \bibfield  {author} {\bibinfo {author} {\bibfnamefont {A.~K.}\ \bibnamefont
  {Ekert}},\ }\href {\doibase 10.1103/PhysRevLett.67.661} {\bibfield  {journal}
  {\bibinfo  {journal} {Phys. Rev. Lett.}\ }\textbf {\bibinfo {volume} {67}},\
  \bibinfo {pages} {661} (\bibinfo {year} {1991})}\BibitemShut {NoStop}%
\bibitem [{\citenamefont {Bennett}\ and\ \citenamefont
  {Wiesner}(1992)}]{bennett1992communication}%
  \BibitemOpen
  \bibfield  {author} {\bibinfo {author} {\bibfnamefont {C.~H.}\ \bibnamefont
  {Bennett}}\ and\ \bibinfo {author} {\bibfnamefont {S.~J.}\ \bibnamefont
  {Wiesner}},\ }\href {\doibase 10.1103/PhysRevLett.69.2881} {\bibfield
  {journal} {\bibinfo  {journal} {Phys. Rev. Lett.}\ }\textbf {\bibinfo
  {volume} {69}},\ \bibinfo {pages} {2881} (\bibinfo {year}
  {1992})}\BibitemShut {NoStop}%
\bibitem [{\citenamefont {Nielsen}\ and\ \citenamefont
  {Chuang}(2002)}]{nielsen2002quantum}%
  \BibitemOpen
  \bibfield  {author} {\bibinfo {author} {\bibfnamefont {M.~A.}\ \bibnamefont
  {Nielsen}}\ and\ \bibinfo {author} {\bibfnamefont {I.}~\bibnamefont
  {Chuang}},\ }\href@noop {} {\  (\bibinfo {year} {2002})}\BibitemShut
  {NoStop}%
\bibitem [{\citenamefont {Wiseman}\ and\ \citenamefont
  {Vaccaro}(2003)}]{wiseman2003entanglement}%
  \BibitemOpen
  \bibfield  {author} {\bibinfo {author} {\bibfnamefont {H.}~\bibnamefont
  {Wiseman}}\ and\ \bibinfo {author} {\bibfnamefont {J.~A.}\ \bibnamefont
  {Vaccaro}},\ }\href@noop {} {\bibfield  {journal} {\bibinfo  {journal} {Phys.
  Rev. Lett.}\ }\textbf {\bibinfo {volume} {91}},\ \bibinfo {pages} {097902}
  (\bibinfo {year} {2003})}\BibitemShut {NoStop}%
\bibitem [{\citenamefont {Barghathi}\ \emph
  {et~al.}(2019{\natexlab{a}})\citenamefont {Barghathi}, \citenamefont
  {Casiano-Diaz},\ and\ \citenamefont {Del~Maestro}}]{PhysRevA.100.022324}%
  \BibitemOpen
  \bibfield  {author} {\bibinfo {author} {\bibfnamefont {H.}~\bibnamefont
  {Barghathi}}, \bibinfo {author} {\bibfnamefont {E.}~\bibnamefont
  {Casiano-Diaz}}, \ and\ \bibinfo {author} {\bibfnamefont {A.}~\bibnamefont
  {Del~Maestro}},\ }\href {\doibase 10.1103/PhysRevA.100.022324} {\bibfield
  {journal} {\bibinfo  {journal} {Phys. Rev. A}\ }\textbf {\bibinfo {volume}
  {100}},\ \bibinfo {pages} {022324} (\bibinfo {year}
  {2019}{\natexlab{a}})}\BibitemShut {NoStop}%
\bibitem [{\citenamefont {Barghathi}\ \emph
  {et~al.}(2019{\natexlab{b}})\citenamefont {Barghathi}, \citenamefont
  {Casiano-Diaz},\ and\ \citenamefont
  {Del~Maestro}}]{barghathi2019operationally}%
  \BibitemOpen
  \bibfield  {author} {\bibinfo {author} {\bibfnamefont {H.}~\bibnamefont
  {Barghathi}}, \bibinfo {author} {\bibfnamefont {E.}~\bibnamefont
  {Casiano-Diaz}}, \ and\ \bibinfo {author} {\bibfnamefont {A.}~\bibnamefont
  {Del~Maestro}},\ }\href@noop {} {\bibfield  {journal} {\bibinfo  {journal}
  {Phys. Rev. A}\ }\textbf {\bibinfo {volume} {100}},\ \bibinfo {pages}
  {022324} (\bibinfo {year} {2019}{\natexlab{b}})}\BibitemShut {NoStop}%
\bibitem [{\citenamefont {Klich}\ and\ \citenamefont
  {Levitov}(2008)}]{klich2008scaling}%
  \BibitemOpen
  \bibfield  {author} {\bibinfo {author} {\bibfnamefont {I.}~\bibnamefont
  {Klich}}\ and\ \bibinfo {author} {\bibfnamefont {L.}~\bibnamefont
  {Levitov}},\ }\href@noop {} {\bibfield  {journal} {\bibinfo  {journal}
  {arXiv:0812.0006}\ } (\bibinfo {year} {2008})}\BibitemShut {NoStop}%
\bibitem [{\citenamefont {Barghathi}\ \emph {et~al.}(2018)\citenamefont
  {Barghathi}, \citenamefont {Herdman},\ and\ \citenamefont
  {Del~Maestro}}]{barghathi2018renyi}%
  \BibitemOpen
  \bibfield  {author} {\bibinfo {author} {\bibfnamefont {H.}~\bibnamefont
  {Barghathi}}, \bibinfo {author} {\bibfnamefont {C.}~\bibnamefont {Herdman}},
  \ and\ \bibinfo {author} {\bibfnamefont {A.}~\bibnamefont {Del~Maestro}},\
  }\href@noop {} {\bibfield  {journal} {\bibinfo  {journal} {Phys. Rev. Lett.}\
  }\textbf {\bibinfo {volume} {121}},\ \bibinfo {pages} {150501} (\bibinfo
  {year} {2018})}\BibitemShut {NoStop}%
\bibitem [{\citenamefont {Goldstein}\ and\ \citenamefont
  {Sela}(2018)}]{goldstein2018symmetry}%
  \BibitemOpen
  \bibfield  {author} {\bibinfo {author} {\bibfnamefont {M.}~\bibnamefont
  {Goldstein}}\ and\ \bibinfo {author} {\bibfnamefont {E.}~\bibnamefont
  {Sela}},\ }\href@noop {} {\bibfield  {journal} {\bibinfo  {journal} {Phys.
  Rev. Lett.}\ }\textbf {\bibinfo {volume} {120}},\ \bibinfo {pages} {200602}
  (\bibinfo {year} {2018})}\BibitemShut {NoStop}%
\bibitem [{\citenamefont {Xavier}\ \emph {et~al.}(2018)\citenamefont {Xavier},
  \citenamefont {Alcaraz},\ and\ \citenamefont
  {Sierra}}]{xavier2018equipartition}%
  \BibitemOpen
  \bibfield  {author} {\bibinfo {author} {\bibfnamefont {J.}~\bibnamefont
  {Xavier}}, \bibinfo {author} {\bibfnamefont {F.~C.}\ \bibnamefont {Alcaraz}},
  \ and\ \bibinfo {author} {\bibfnamefont {G.}~\bibnamefont {Sierra}},\
  }\href@noop {} {\bibfield  {journal} {\bibinfo  {journal} {Phys. Rev. B}\
  }\textbf {\bibinfo {volume} {98}},\ \bibinfo {pages} {041106} (\bibinfo
  {year} {2018})}\BibitemShut {NoStop}%
\bibitem [{\citenamefont {Vaccaro}\ \emph {et~al.}(2008)\citenamefont
  {Vaccaro}, \citenamefont {Anselmi}, \citenamefont {Wiseman},\ and\
  \citenamefont {Jacobs}}]{vaccaro2008tradeoff}%
  \BibitemOpen
  \bibfield  {author} {\bibinfo {author} {\bibfnamefont {J.~A.}\ \bibnamefont
  {Vaccaro}}, \bibinfo {author} {\bibfnamefont {F.}~\bibnamefont {Anselmi}},
  \bibinfo {author} {\bibfnamefont {H.~M.}\ \bibnamefont {Wiseman}}, \ and\
  \bibinfo {author} {\bibfnamefont {K.}~\bibnamefont {Jacobs}},\ }\href@noop {}
  {\bibfield  {journal} {\bibinfo  {journal} {Phys. Rev. A}\ }\textbf {\bibinfo
  {volume} {77}},\ \bibinfo {pages} {032114} (\bibinfo {year}
  {2008})}\BibitemShut {NoStop}%
\bibitem [{\citenamefont {Wick}\ \emph {et~al.}(1970)\citenamefont {Wick},
  \citenamefont {Wightman},\ and\ \citenamefont
  {Wigner}}]{wick1970superselection}%
  \BibitemOpen
  \bibfield  {author} {\bibinfo {author} {\bibfnamefont {G.-C.}\ \bibnamefont
  {Wick}}, \bibinfo {author} {\bibfnamefont {A.~S.}\ \bibnamefont {Wightman}},
  \ and\ \bibinfo {author} {\bibfnamefont {E.~P.}\ \bibnamefont {Wigner}},\
  }\href@noop {} {\bibfield  {journal} {\bibinfo  {journal} {Phys. Rev. D}\
  }\textbf {\bibinfo {volume} {1}},\ \bibinfo {pages} {3267} (\bibinfo {year}
  {1970})}\BibitemShut {NoStop}%
\bibitem [{\citenamefont {Schuch}\ \emph {et~al.}(2004)\citenamefont {Schuch},
  \citenamefont {Verstraete},\ and\ \citenamefont
  {Cirac}}]{PhysRevA.70.042310}%
  \BibitemOpen
  \bibfield  {author} {\bibinfo {author} {\bibfnamefont {N.}~\bibnamefont
  {Schuch}}, \bibinfo {author} {\bibfnamefont {F.}~\bibnamefont {Verstraete}},
  \ and\ \bibinfo {author} {\bibfnamefont {J.~I.}\ \bibnamefont {Cirac}},\
  }\href {\doibase 10.1103/PhysRevA.70.042310} {\bibfield  {journal} {\bibinfo
  {journal} {Phys. Rev. A}\ }\textbf {\bibinfo {volume} {70}},\ \bibinfo
  {pages} {042310} (\bibinfo {year} {2004})}\BibitemShut {NoStop}%
\bibitem [{\citenamefont {Verstraete}\ and\ \citenamefont
  {Cirac}(2003)}]{verstraete2003quantum}%
  \BibitemOpen
  \bibfield  {author} {\bibinfo {author} {\bibfnamefont {F.}~\bibnamefont
  {Verstraete}}\ and\ \bibinfo {author} {\bibfnamefont {J.~I.}\ \bibnamefont
  {Cirac}},\ }\href@noop {} {\bibfield  {journal} {\bibinfo  {journal} {Phys.
  Rev. Lett.}\ }\textbf {\bibinfo {volume} {91}},\ \bibinfo {pages} {010404}
  (\bibinfo {year} {2003})}\BibitemShut {NoStop}%
\bibitem [{\citenamefont {Azses}\ \emph {et~al.}(2020)\citenamefont {Azses},
  \citenamefont {Haenel}, \citenamefont {Naveh}, \citenamefont {Raussendorf},
  \citenamefont {Sela},\ and\ \citenamefont {Torre}}]{2002.04620}%
  \BibitemOpen
  \bibfield  {author} {\bibinfo {author} {\bibfnamefont {D.}~\bibnamefont
  {Azses}}, \bibinfo {author} {\bibfnamefont {R.}~\bibnamefont {Haenel}},
  \bibinfo {author} {\bibfnamefont {Y.}~\bibnamefont {Naveh}}, \bibinfo
  {author} {\bibfnamefont {R.}~\bibnamefont {Raussendorf}}, \bibinfo {author}
  {\bibfnamefont {E.}~\bibnamefont {Sela}}, \ and\ \bibinfo {author}
  {\bibfnamefont {E.~G.~D.}\ \bibnamefont {Torre}},\ }\href@noop {} {\
  (\bibinfo {year} {2020})},\ \Eprint {http://arxiv.org/abs/arXiv:2002.04620}
  {arXiv:2002.04620} \BibitemShut {NoStop}%
\bibitem [{\citenamefont {Fraenkel}\ and\ \citenamefont
  {Goldstein}(2019)}]{fraenkel2019symmetry}%
  \BibitemOpen
  \bibfield  {author} {\bibinfo {author} {\bibfnamefont {S.}~\bibnamefont
  {Fraenkel}}\ and\ \bibinfo {author} {\bibfnamefont {M.}~\bibnamefont
  {Goldstein}},\ }\href@noop {} {\bibfield  {journal} {\bibinfo  {journal}
  {arXiv:1910.08459}\ } (\bibinfo {year} {2019})}\BibitemShut {NoStop}%
\bibitem [{\citenamefont {Calabrese}\ \emph {et~al.}(2020)\citenamefont
  {Calabrese}, \citenamefont {Collura}, \citenamefont {Giulio},\ and\
  \citenamefont {Murciano}}]{2002.04367}%
  \BibitemOpen
  \bibfield  {author} {\bibinfo {author} {\bibfnamefont {P.}~\bibnamefont
  {Calabrese}}, \bibinfo {author} {\bibfnamefont {M.}~\bibnamefont {Collura}},
  \bibinfo {author} {\bibfnamefont {G.~D.}\ \bibnamefont {Giulio}}, \ and\
  \bibinfo {author} {\bibfnamefont {S.}~\bibnamefont {Murciano}},\ }\href@noop
  {} {\  (\bibinfo {year} {2020})},\ \Eprint
  {http://arxiv.org/abs/arXiv:2002.04367} {arXiv:2002.04367} \BibitemShut
  {NoStop}%
\bibitem [{\citenamefont {Cornfeld}\ \emph {et~al.}(2019)\citenamefont
  {Cornfeld}, \citenamefont {Landau}, \citenamefont {Shtengel},\ and\
  \citenamefont {Sela}}]{cornfeld2019entanglement}%
  \BibitemOpen
  \bibfield  {author} {\bibinfo {author} {\bibfnamefont {E.}~\bibnamefont
  {Cornfeld}}, \bibinfo {author} {\bibfnamefont {L.~A.}\ \bibnamefont
  {Landau}}, \bibinfo {author} {\bibfnamefont {K.}~\bibnamefont {Shtengel}}, \
  and\ \bibinfo {author} {\bibfnamefont {E.}~\bibnamefont {Sela}},\ }\href@noop
  {} {\bibfield  {journal} {\bibinfo  {journal} {Phys. Rev. B}\ }\textbf
  {\bibinfo {volume} {99}},\ \bibinfo {pages} {115429} (\bibinfo {year}
  {2019})}\BibitemShut {NoStop}%
\bibitem [{\citenamefont {Bartlett}\ and\ \citenamefont
  {Wiseman}(2003)}]{bartlett2003entanglement}%
  \BibitemOpen
  \bibfield  {author} {\bibinfo {author} {\bibfnamefont {S.~D.}\ \bibnamefont
  {Bartlett}}\ and\ \bibinfo {author} {\bibfnamefont {H.~M.}\ \bibnamefont
  {Wiseman}},\ }\href@noop {} {\bibfield  {journal} {\bibinfo  {journal} {Phys.
  Rev. Lett.}\ }\textbf {\bibinfo {volume} {91}},\ \bibinfo {pages} {097903}
  (\bibinfo {year} {2003})}\BibitemShut {NoStop}%
\bibitem [{\citenamefont {Murciano}\ \emph {et~al.}(2019)\citenamefont
  {Murciano}, \citenamefont {Giulio},\ and\ \citenamefont
  {Calabrese}}]{1911.09588}%
  \BibitemOpen
  \bibfield  {author} {\bibinfo {author} {\bibfnamefont {S.}~\bibnamefont
  {Murciano}}, \bibinfo {author} {\bibfnamefont {G.~D.}\ \bibnamefont
  {Giulio}}, \ and\ \bibinfo {author} {\bibfnamefont {P.}~\bibnamefont
  {Calabrese}},\ }\href@noop {} {\  (\bibinfo {year} {2019})},\ \Eprint
  {http://arxiv.org/abs/arXiv:1911.09588} {arXiv:1911.09588} \BibitemShut
  {NoStop}%
\bibitem [{\citenamefont {Laflorencie}\ and\ \citenamefont
  {Rachel}(2014)}]{laflorencie2014spin}%
  \BibitemOpen
  \bibfield  {author} {\bibinfo {author} {\bibfnamefont {N.}~\bibnamefont
  {Laflorencie}}\ and\ \bibinfo {author} {\bibfnamefont {S.}~\bibnamefont
  {Rachel}},\ }\href@noop {} {\bibfield  {journal} {\bibinfo  {journal}
  {Journal of Statistical Mechanics: Theory and Experiment}\ }\textbf {\bibinfo
  {volume} {2014}},\ \bibinfo {pages} {P11013} (\bibinfo {year}
  {2014})}\BibitemShut {NoStop}%
\bibitem [{\citenamefont {Ghosh}\ \emph {et~al.}(2015)\citenamefont {Ghosh},
  \citenamefont {Soni},\ and\ \citenamefont {Trivedi}}]{ghosh2015entanglement}%
  \BibitemOpen
  \bibfield  {author} {\bibinfo {author} {\bibfnamefont {S.}~\bibnamefont
  {Ghosh}}, \bibinfo {author} {\bibfnamefont {R.~M.}\ \bibnamefont {Soni}}, \
  and\ \bibinfo {author} {\bibfnamefont {S.~P.}\ \bibnamefont {Trivedi}},\
  }\href@noop {} {\bibfield  {journal} {\bibinfo  {journal} {Journal of High
  Energy Physics}\ }\textbf {\bibinfo {volume} {2015}},\ \bibinfo {pages} {69}
  (\bibinfo {year} {2015})}\BibitemShut {NoStop}%
\bibitem [{\citenamefont {Van~Acoleyen}\ \emph {et~al.}(2016)\citenamefont
  {Van~Acoleyen}, \citenamefont {Bultinck}, \citenamefont {Haegeman},
  \citenamefont {Marien}, \citenamefont {Scholz},\ and\ \citenamefont
  {Verstraete}}]{PhysRevLett.117.131602}%
  \BibitemOpen
  \bibfield  {author} {\bibinfo {author} {\bibfnamefont {K.}~\bibnamefont
  {Van~Acoleyen}}, \bibinfo {author} {\bibfnamefont {N.}~\bibnamefont
  {Bultinck}}, \bibinfo {author} {\bibfnamefont {J.}~\bibnamefont {Haegeman}},
  \bibinfo {author} {\bibfnamefont {M.}~\bibnamefont {Marien}}, \bibinfo
  {author} {\bibfnamefont {V.~B.}\ \bibnamefont {Scholz}}, \ and\ \bibinfo
  {author} {\bibfnamefont {F.}~\bibnamefont {Verstraete}},\ }\href {\doibase
  10.1103/PhysRevLett.117.131602} {\bibfield  {journal} {\bibinfo  {journal}
  {Phys. Rev. Lett.}\ }\textbf {\bibinfo {volume} {117}},\ \bibinfo {pages}
  {131602} (\bibinfo {year} {2016})}\BibitemShut {NoStop}%
\bibitem [{\citenamefont {Soni}\ and\ \citenamefont
  {Trivedi}(2016)}]{soni2016aspects}%
  \BibitemOpen
  \bibfield  {author} {\bibinfo {author} {\bibfnamefont {R.~M.}\ \bibnamefont
  {Soni}}\ and\ \bibinfo {author} {\bibfnamefont {S.~P.}\ \bibnamefont
  {Trivedi}},\ }\href@noop {} {\bibfield  {journal} {\bibinfo  {journal}
  {Journal of High Energy Physics}\ }\textbf {\bibinfo {volume} {2016}},\
  \bibinfo {pages} {136} (\bibinfo {year} {2016})}\BibitemShut {NoStop}%
\bibitem [{\citenamefont {Chen}\ \emph {et~al.}(2010)\citenamefont {Chen},
  \citenamefont {Gu},\ and\ \citenamefont {Wen}}]{chen2010local}%
  \BibitemOpen
  \bibfield  {author} {\bibinfo {author} {\bibfnamefont {X.}~\bibnamefont
  {Chen}}, \bibinfo {author} {\bibfnamefont {Z.-C.}\ \bibnamefont {Gu}}, \ and\
  \bibinfo {author} {\bibfnamefont {X.-G.}\ \bibnamefont {Wen}},\ }\href@noop
  {} {\bibfield  {journal} {\bibinfo  {journal} {Phys. Rev. B}\ }\textbf
  {\bibinfo {volume} {82}},\ \bibinfo {pages} {155138} (\bibinfo {year}
  {2010})}\BibitemShut {NoStop}%
\bibitem [{\citenamefont {Chen}\ \emph {et~al.}(2013)\citenamefont {Chen},
  \citenamefont {Gu}, \citenamefont {Liu},\ and\ \citenamefont
  {Wen}}]{prevWork3}%
  \BibitemOpen
  \bibfield  {author} {\bibinfo {author} {\bibfnamefont {X.}~\bibnamefont
  {Chen}}, \bibinfo {author} {\bibfnamefont {Z.-C.}\ \bibnamefont {Gu}},
  \bibinfo {author} {\bibfnamefont {Z.-X.}\ \bibnamefont {Liu}}, \ and\
  \bibinfo {author} {\bibfnamefont {X.-G.}\ \bibnamefont {Wen}},\ }\href@noop
  {} {\bibfield  {journal} {\bibinfo  {journal} {Phys. Rev. B}\ }\textbf
  {\bibinfo {volume} {87}},\ \bibinfo {pages} {155114} (\bibinfo {year}
  {2013})}\BibitemShut {NoStop}%
\bibitem [{\citenamefont {Wen}(2017)}]{wen2017colloquium}%
  \BibitemOpen
  \bibfield  {author} {\bibinfo {author} {\bibfnamefont {X.-G.}\ \bibnamefont
  {Wen}},\ }\href@noop {} {\bibfield  {journal} {\bibinfo  {journal} {Reviews
  of Modern Physics}\ }\textbf {\bibinfo {volume} {89}},\ \bibinfo {pages}
  {041004} (\bibinfo {year} {2017})}\BibitemShut {NoStop}%
\bibitem [{\citenamefont {Pollmann}\ \emph {et~al.}(2010)\citenamefont
  {Pollmann}, \citenamefont {Turner}, \citenamefont {Berg},\ and\ \citenamefont
  {Oshikawa}}]{pollmann2010}%
  \BibitemOpen
  \bibfield  {author} {\bibinfo {author} {\bibfnamefont {F.}~\bibnamefont
  {Pollmann}}, \bibinfo {author} {\bibfnamefont {A.~M.}\ \bibnamefont
  {Turner}}, \bibinfo {author} {\bibfnamefont {E.}~\bibnamefont {Berg}}, \ and\
  \bibinfo {author} {\bibfnamefont {M.}~\bibnamefont {Oshikawa}},\ }\href
  {\doibase 10.1103/PhysRevB.81.064439} {\bibfield  {journal} {\bibinfo
  {journal} {Phys. Rev. B}\ }\textbf {\bibinfo {volume} {81}},\ \bibinfo
  {pages} {064439} (\bibinfo {year} {2010})}\BibitemShut {NoStop}%
\bibitem [{\citenamefont {Else}\ and\ \citenamefont
  {Nayak}(2014)}]{else2014classifying}%
  \BibitemOpen
  \bibfield  {author} {\bibinfo {author} {\bibfnamefont {D.~V.}\ \bibnamefont
  {Else}}\ and\ \bibinfo {author} {\bibfnamefont {C.}~\bibnamefont {Nayak}},\
  }\href {\doibase 10.1103/PhysRevB.90.235137} {\bibfield  {journal} {\bibinfo
  {journal} {Phys. Rev. B}\ }\textbf {\bibinfo {volume} {90}},\ \bibinfo
  {pages} {235137} (\bibinfo {year} {2014})}\BibitemShut {NoStop}%
\bibitem [{\citenamefont {Schuch}\ \emph {et~al.}(2011)\citenamefont {Schuch},
  \citenamefont {P{\'e}rez-Garc{\'\i}a},\ and\ \citenamefont
  {Cirac}}]{schuch2011classifying}%
  \BibitemOpen
  \bibfield  {author} {\bibinfo {author} {\bibfnamefont {N.}~\bibnamefont
  {Schuch}}, \bibinfo {author} {\bibfnamefont {D.}~\bibnamefont
  {P{\'e}rez-Garc{\'\i}a}}, \ and\ \bibinfo {author} {\bibfnamefont
  {I.}~\bibnamefont {Cirac}},\ }\href@noop {} {\bibfield  {journal} {\bibinfo
  {journal} {Phys. Rev. B}\ }\textbf {\bibinfo {volume} {84}},\ \bibinfo
  {pages} {165139} (\bibinfo {year} {2011})}\BibitemShut {NoStop}%
\bibitem [{\citenamefont {Chen}\ \emph {et~al.}(2011)\citenamefont {Chen},
  \citenamefont {Gu},\ and\ \citenamefont {Wen}}]{chen2011classification}%
  \BibitemOpen
  \bibfield  {author} {\bibinfo {author} {\bibfnamefont {X.}~\bibnamefont
  {Chen}}, \bibinfo {author} {\bibfnamefont {Z.-C.}\ \bibnamefont {Gu}}, \ and\
  \bibinfo {author} {\bibfnamefont {X.-G.}\ \bibnamefont {Wen}},\ }\href@noop
  {} {\bibfield  {journal} {\bibinfo  {journal} {Phys. Rev. B}\ }\textbf
  {\bibinfo {volume} {83}},\ \bibinfo {pages} {035107} (\bibinfo {year}
  {2011})}\BibitemShut {NoStop}%
\bibitem [{\citenamefont {Wen}(2019)}]{wen2019choreographed}%
  \BibitemOpen
  \bibfield  {author} {\bibinfo {author} {\bibfnamefont {X.-G.}\ \bibnamefont
  {Wen}},\ }\href@noop {} {\bibfield  {journal} {\bibinfo  {journal}
  {arXiv:1906.05983}\ } (\bibinfo {year} {2019})}\BibitemShut {NoStop}%
\bibitem [{\citenamefont {Zeng}\ \emph {et~al.}(2015)\citenamefont {Zeng},
  \citenamefont {Chen}, \citenamefont {Zhou},\ and\ \citenamefont
  {Wen}}]{prevWork2}%
  \BibitemOpen
  \bibfield  {author} {\bibinfo {author} {\bibfnamefont {B.}~\bibnamefont
  {Zeng}}, \bibinfo {author} {\bibfnamefont {X.}~\bibnamefont {Chen}}, \bibinfo
  {author} {\bibfnamefont {D.-L.}\ \bibnamefont {Zhou}}, \ and\ \bibinfo
  {author} {\bibfnamefont {X.-G.}\ \bibnamefont {Wen}},\ }\href@noop {}
  {\bibfield  {journal} {\bibinfo  {journal} {arXiv:1508.02595}\ } (\bibinfo
  {year} {2015})}\BibitemShut {NoStop}%
\bibitem [{\citenamefont {Duivenvoorden}\ \emph {et~al.}(2017)\citenamefont
  {Duivenvoorden}, \citenamefont {Iqbal}, \citenamefont {Haegeman},
  \citenamefont {Verstraete},\ and\ \citenamefont
  {Schuch}}]{duivenvoorden2017entanglement}%
  \BibitemOpen
  \bibfield  {author} {\bibinfo {author} {\bibfnamefont {K.}~\bibnamefont
  {Duivenvoorden}}, \bibinfo {author} {\bibfnamefont {M.}~\bibnamefont
  {Iqbal}}, \bibinfo {author} {\bibfnamefont {J.}~\bibnamefont {Haegeman}},
  \bibinfo {author} {\bibfnamefont {F.}~\bibnamefont {Verstraete}}, \ and\
  \bibinfo {author} {\bibfnamefont {N.}~\bibnamefont {Schuch}},\ }\href@noop {}
  {\bibfield  {journal} {\bibinfo  {journal} {Phys. Rev. B}\ }\textbf {\bibinfo
  {volume} {95}},\ \bibinfo {pages} {235119} (\bibinfo {year}
  {2017})}\BibitemShut {NoStop}%
\bibitem [{\citenamefont {Verresen}\ \emph {et~al.}(2017)\citenamefont
  {Verresen}, \citenamefont {Moessner},\ and\ \citenamefont
  {Pollmann}}]{verresen2017one}%
  \BibitemOpen
  \bibfield  {author} {\bibinfo {author} {\bibfnamefont {R.}~\bibnamefont
  {Verresen}}, \bibinfo {author} {\bibfnamefont {R.}~\bibnamefont {Moessner}},
  \ and\ \bibinfo {author} {\bibfnamefont {F.}~\bibnamefont {Pollmann}},\
  }\href@noop {} {\bibfield  {journal} {\bibinfo  {journal} {Phys. Rev. B}\
  }\textbf {\bibinfo {volume} {96}},\ \bibinfo {pages} {165124} (\bibinfo
  {year} {2017})}\BibitemShut {NoStop}%
\bibitem [{\citenamefont {Wen}(2002)}]{wen2002quantum}%
  \BibitemOpen
  \bibfield  {author} {\bibinfo {author} {\bibfnamefont {X.-G.}\ \bibnamefont
  {Wen}},\ }\href@noop {} {\bibfield  {journal} {\bibinfo  {journal} {Physics
  Letters A}\ }\textbf {\bibinfo {volume} {300}},\ \bibinfo {pages} {175}
  (\bibinfo {year} {2002})}\BibitemShut {NoStop}%
\bibitem [{\citenamefont {Marvian}(2017)}]{iman}%
  \BibitemOpen
  \bibfield  {author} {\bibinfo {author} {\bibfnamefont {I.}~\bibnamefont
  {Marvian}},\ }\href@noop {} {\bibfield  {journal} {\bibinfo  {journal} {Phys.
  Rev. B}\ }\textbf {\bibinfo {volume} {95}},\ \bibinfo {pages} {045111}
  (\bibinfo {year} {2017})}\BibitemShut {NoStop}%
\bibitem [{\citenamefont {Pollmann}\ and\ \citenamefont
  {Turner}(2012)}]{pollmann2012detection}%
  \BibitemOpen
  \bibfield  {author} {\bibinfo {author} {\bibfnamefont {F.}~\bibnamefont
  {Pollmann}}\ and\ \bibinfo {author} {\bibfnamefont {A.~M.}\ \bibnamefont
  {Turner}},\ }\href@noop {} {\bibfield  {journal} {\bibinfo  {journal} {Phys.
  Rev. B}\ }\textbf {\bibinfo {volume} {86}},\ \bibinfo {pages} {125441}
  (\bibinfo {year} {2012})}\BibitemShut {NoStop}%
\bibitem [{\citenamefont {den Nijs}\ and\ \citenamefont
  {Rommelse}(1989{\natexlab{a}})}]{PhysRevB.40.4709}%
  \BibitemOpen
  \bibfield  {author} {\bibinfo {author} {\bibfnamefont {M.}~\bibnamefont {den
  Nijs}}\ and\ \bibinfo {author} {\bibfnamefont {K.}~\bibnamefont {Rommelse}},\
  }\href {\doibase 10.1103/PhysRevB.40.4709} {\bibfield  {journal} {\bibinfo
  {journal} {Phys. Rev. B}\ }\textbf {\bibinfo {volume} {40}},\ \bibinfo
  {pages} {4709} (\bibinfo {year} {1989}{\natexlab{a}})}\BibitemShut {NoStop}%
\bibitem [{\citenamefont {Haegeman}\ \emph {et~al.}(2012)\citenamefont
  {Haegeman}, \citenamefont {P{\'e}rez-Garc{\'\i}a}, \citenamefont {Cirac},\
  and\ \citenamefont {Schuch}}]{haegeman2012order}%
  \BibitemOpen
  \bibfield  {author} {\bibinfo {author} {\bibfnamefont {J.}~\bibnamefont
  {Haegeman}}, \bibinfo {author} {\bibfnamefont {D.}~\bibnamefont
  {P{\'e}rez-Garc{\'\i}a}}, \bibinfo {author} {\bibfnamefont {I.}~\bibnamefont
  {Cirac}}, \ and\ \bibinfo {author} {\bibfnamefont {N.}~\bibnamefont
  {Schuch}},\ }\href@noop {} {\bibfield  {journal} {\bibinfo  {journal} {Phys.
  Rev. Lett.}\ }\textbf {\bibinfo {volume} {109}},\ \bibinfo {pages} {050402}
  (\bibinfo {year} {2012})}\BibitemShut {NoStop}%
\bibitem [{\citenamefont {Stephen}\ \emph
  {et~al.}(2019{\natexlab{a}})\citenamefont {Stephen}, \citenamefont {Dreyer},
  \citenamefont {Iqbal},\ and\ \citenamefont {Schuch}}]{stephen2019detecting}%
  \BibitemOpen
  \bibfield  {author} {\bibinfo {author} {\bibfnamefont {D.~T.}\ \bibnamefont
  {Stephen}}, \bibinfo {author} {\bibfnamefont {H.}~\bibnamefont {Dreyer}},
  \bibinfo {author} {\bibfnamefont {M.}~\bibnamefont {Iqbal}}, \ and\ \bibinfo
  {author} {\bibfnamefont {N.}~\bibnamefont {Schuch}},\ }\href {\doibase
  10.1103/PhysRevB.100.115112} {\bibfield  {journal} {\bibinfo  {journal}
  {Phys. Rev. B}\ }\textbf {\bibinfo {volume} {100}},\ \bibinfo {pages}
  {115112} (\bibinfo {year} {2019}{\natexlab{a}})}\BibitemShut {NoStop}%
\bibitem [{\citenamefont {Else}\ \emph
  {et~al.}(2012{\natexlab{a}})\citenamefont {Else}, \citenamefont {Schwarz},
  \citenamefont {Bartlett},\ and\ \citenamefont {Doherty}}]{prevWork1}%
  \BibitemOpen
  \bibfield  {author} {\bibinfo {author} {\bibfnamefont {D.~V.}\ \bibnamefont
  {Else}}, \bibinfo {author} {\bibfnamefont {I.}~\bibnamefont {Schwarz}},
  \bibinfo {author} {\bibfnamefont {S.~D.}\ \bibnamefont {Bartlett}}, \ and\
  \bibinfo {author} {\bibfnamefont {A.~C.}\ \bibnamefont {Doherty}},\ }\href
  {\doibase 10.1103/PhysRevLett.108.240505} {\bibfield  {journal} {\bibinfo
  {journal} {Phys. Rev. Lett.}\ }\textbf {\bibinfo {volume} {108}},\ \bibinfo
  {pages} {240505} (\bibinfo {year} {2012}{\natexlab{a}})}\BibitemShut
  {NoStop}%
\bibitem [{\citenamefont {You}\ \emph {et~al.}(2018)\citenamefont {You},
  \citenamefont {Devakul}, \citenamefont {Burnell},\ and\ \citenamefont
  {Sondhi}}]{you2018subsystem}%
  \BibitemOpen
  \bibfield  {author} {\bibinfo {author} {\bibfnamefont {Y.}~\bibnamefont
  {You}}, \bibinfo {author} {\bibfnamefont {T.}~\bibnamefont {Devakul}},
  \bibinfo {author} {\bibfnamefont {F.~J.}\ \bibnamefont {Burnell}}, \ and\
  \bibinfo {author} {\bibfnamefont {S.~L.}\ \bibnamefont {Sondhi}},\
  }\href@noop {} {\bibfield  {journal} {\bibinfo  {journal} {Phys. Rev. B}\
  }\textbf {\bibinfo {volume} {98}},\ \bibinfo {pages} {035112} (\bibinfo
  {year} {2018})}\BibitemShut {NoStop}%
\bibitem [{\citenamefont {Stephen}\ \emph
  {et~al.}(2019{\natexlab{b}})\citenamefont {Stephen}, \citenamefont {Nautrup},
  \citenamefont {Bermejo-Vega}, \citenamefont {Eisert},\ and\ \citenamefont
  {Raussendorf}}]{stephen2019subsystem}%
  \BibitemOpen
  \bibfield  {author} {\bibinfo {author} {\bibfnamefont {D.~T.}\ \bibnamefont
  {Stephen}}, \bibinfo {author} {\bibfnamefont {H.~P.}\ \bibnamefont
  {Nautrup}}, \bibinfo {author} {\bibfnamefont {J.}~\bibnamefont
  {Bermejo-Vega}}, \bibinfo {author} {\bibfnamefont {J.}~\bibnamefont
  {Eisert}}, \ and\ \bibinfo {author} {\bibfnamefont {R.}~\bibnamefont
  {Raussendorf}},\ }\href@noop {} {\bibfield  {journal} {\bibinfo  {journal}
  {Quantum}\ }\textbf {\bibinfo {volume} {3}},\ \bibinfo {pages} {142}
  (\bibinfo {year} {2019}{\natexlab{b}})}\BibitemShut {NoStop}%
\bibitem [{\citenamefont {Raussendorf}\ \emph {et~al.}(2003)\citenamefont
  {Raussendorf}, \citenamefont {Browne},\ and\ \citenamefont
  {Briegel}}]{raussendorf2003measurement}%
  \BibitemOpen
  \bibfield  {author} {\bibinfo {author} {\bibfnamefont {R.}~\bibnamefont
  {Raussendorf}}, \bibinfo {author} {\bibfnamefont {D.~E.}\ \bibnamefont
  {Browne}}, \ and\ \bibinfo {author} {\bibfnamefont {H.~J.}\ \bibnamefont
  {Briegel}},\ }\href@noop {} {\bibfield  {journal} {\bibinfo  {journal} {Phys.
  Rev. A}\ }\textbf {\bibinfo {volume} {68}},\ \bibinfo {pages} {022312}
  (\bibinfo {year} {2003})}\BibitemShut {NoStop}%
\bibitem [{\citenamefont {Stephen}\ \emph {et~al.}(2017)\citenamefont
  {Stephen}, \citenamefont {Wang}, \citenamefont {Prakash}, \citenamefont
  {Wei},\ and\ \citenamefont {Raussendorf}}]{stephen2017computational}%
  \BibitemOpen
  \bibfield  {author} {\bibinfo {author} {\bibfnamefont {D.~T.}\ \bibnamefont
  {Stephen}}, \bibinfo {author} {\bibfnamefont {D.-S.}\ \bibnamefont {Wang}},
  \bibinfo {author} {\bibfnamefont {A.}~\bibnamefont {Prakash}}, \bibinfo
  {author} {\bibfnamefont {T.-C.}\ \bibnamefont {Wei}}, \ and\ \bibinfo
  {author} {\bibfnamefont {R.}~\bibnamefont {Raussendorf}},\ }\href {\doibase
  10.1103/PhysRevLett.119.010504} {\bibfield  {journal} {\bibinfo  {journal}
  {Phys. Rev. Lett.}\ }\textbf {\bibinfo {volume} {119}},\ \bibinfo {pages}
  {010504} (\bibinfo {year} {2017})}\BibitemShut {NoStop}%
\bibitem [{\citenamefont {Hastings}(2007)}]{hastings2007area}%
  \BibitemOpen
  \bibfield  {author} {\bibinfo {author} {\bibfnamefont {M.~B.}\ \bibnamefont
  {Hastings}},\ }\href@noop {} {\bibfield  {journal} {\bibinfo  {journal}
  {Journal of Statistical Mechanics: Theory and Experiment}\ }\textbf {\bibinfo
  {volume} {2007}},\ \bibinfo {pages} {P08024} (\bibinfo {year}
  {2007})}\BibitemShut {NoStop}%
\bibitem [{\citenamefont {Verstraete}\ and\ \citenamefont
  {Cirac}(2006)}]{verstraete2006matrix}%
  \BibitemOpen
  \bibfield  {author} {\bibinfo {author} {\bibfnamefont {F.}~\bibnamefont
  {Verstraete}}\ and\ \bibinfo {author} {\bibfnamefont {J.~I.}\ \bibnamefont
  {Cirac}},\ }\href {\doibase 10.1103/PhysRevB.73.094423} {\bibfield  {journal}
  {\bibinfo  {journal} {Phys. Rev. B}\ }\textbf {\bibinfo {volume} {73}},\
  \bibinfo {pages} {094423} (\bibinfo {year} {2006})}\BibitemShut {NoStop}%
\bibitem [{\citenamefont {Perez-Garcia}\ \emph {et~al.}(2007)\citenamefont
  {Perez-Garcia}, \citenamefont {Verstraete}, \citenamefont {Wolf},\ and\
  \citenamefont {Cirac}}]{perez2006matrix}%
  \BibitemOpen
  \bibfield  {author} {\bibinfo {author} {\bibfnamefont {D.}~\bibnamefont
  {Perez-Garcia}}, \bibinfo {author} {\bibfnamefont {F.}~\bibnamefont
  {Verstraete}}, \bibinfo {author} {\bibfnamefont {M.~M.}\ \bibnamefont
  {Wolf}}, \ and\ \bibinfo {author} {\bibfnamefont {J.~I.}\ \bibnamefont
  {Cirac}},\ }\href@noop {} {\bibfield  {journal} {\bibinfo  {journal} {Quantum
  Info. Comput.}\ }\textbf {\bibinfo {volume} {7}},\ \bibinfo {pages}
  {401–430} (\bibinfo {year} {2007})}\BibitemShut {NoStop}%
\bibitem [{\citenamefont {den Nijs}\ and\ \citenamefont
  {Rommelse}(1989{\natexlab{b}})}]{den1989preroughening}%
  \BibitemOpen
  \bibfield  {author} {\bibinfo {author} {\bibfnamefont {M.}~\bibnamefont {den
  Nijs}}\ and\ \bibinfo {author} {\bibfnamefont {K.}~\bibnamefont {Rommelse}},\
  }\href@noop {} {\bibfield  {journal} {\bibinfo  {journal} {Phys. Rev. B}\
  }\textbf {\bibinfo {volume} {40}},\ \bibinfo {pages} {4709} (\bibinfo {year}
  {1989}{\natexlab{b}})}\BibitemShut {NoStop}%
\bibitem [{\citenamefont {Fidkowski}\ \emph {et~al.}(2019)\citenamefont
  {Fidkowski}, \citenamefont {Haah},\ and\ \citenamefont
  {Hastings}}]{1912.05565}%
  \BibitemOpen
  \bibfield  {author} {\bibinfo {author} {\bibfnamefont {L.}~\bibnamefont
  {Fidkowski}}, \bibinfo {author} {\bibfnamefont {J.}~\bibnamefont {Haah}}, \
  and\ \bibinfo {author} {\bibfnamefont {M.~B.}\ \bibnamefont {Hastings}},\
  }\href@noop {} {\  (\bibinfo {year} {2019})},\ \Eprint
  {http://arxiv.org/abs/arXiv:1912.05565} {arXiv:1912.05565} \BibitemShut
  {NoStop}%
\bibitem [{\citenamefont {Karpilovsky}(1994)}]{karpilovsky1994group}%
  \BibitemOpen
  \bibfield  {author} {\bibinfo {author} {\bibfnamefont {G.}~\bibnamefont
  {Karpilovsky}},\ }\href@noop {} {\emph {\bibinfo {title} {Group
  representations}}},\ Vol.~\bibinfo {volume} {3}\ (\bibinfo  {publisher}
  {Elsevier},\ \bibinfo {year} {1994})\BibitemShut {NoStop}%
\bibitem [{\citenamefont {Backhouse}\ and\ \citenamefont
  {Bradley}(1972)}]{backhouse1972projective}%
  \BibitemOpen
  \bibfield  {author} {\bibinfo {author} {\bibfnamefont {N.~B.}\ \bibnamefont
  {Backhouse}}\ and\ \bibinfo {author} {\bibfnamefont {C.}~\bibnamefont
  {Bradley}},\ }\href@noop {} {\bibfield  {journal} {\bibinfo  {journal}
  {Proceedings of the American Mathematical Society}\ }\textbf {\bibinfo
  {volume} {36}},\ \bibinfo {pages} {260} (\bibinfo {year} {1972})}\BibitemShut
  {NoStop}%
\bibitem [{\citenamefont {Berkovich}\ \emph {et~al.}(2018)\citenamefont
  {Berkovich}, \citenamefont {Kazarin},\ and\ \citenamefont
  {Zhmud}}]{berkovich2018yakov}%
  \BibitemOpen
  \bibfield  {author} {\bibinfo {author} {\bibfnamefont {Y.~G.}\ \bibnamefont
  {Berkovich}}, \bibinfo {author} {\bibfnamefont {L.~S.}\ \bibnamefont
  {Kazarin}}, \ and\ \bibinfo {author} {\bibfnamefont {E.~M.}\ \bibnamefont
  {Zhmud}},\ }\href@noop {} {\emph {\bibinfo {title} {Characters of Finite
  Groups}}},\ Vol.~\bibinfo {volume} {2}\ (\bibinfo  {publisher} {Walter de
  Gruyter GmbH \& Co KG},\ \bibinfo {year} {2018})\BibitemShut {NoStop}%
\bibitem [{\citenamefont {Miyake}(2010)}]{miyake2010quantum}%
  \BibitemOpen
  \bibfield  {author} {\bibinfo {author} {\bibfnamefont {A.}~\bibnamefont
  {Miyake}},\ }\href {\doibase 10.1103/PhysRevLett.105.040501} {\bibfield
  {journal} {\bibinfo  {journal} {Phys. Rev. Lett.}\ }\textbf {\bibinfo
  {volume} {105}},\ \bibinfo {pages} {040501} (\bibinfo {year}
  {2010})}\BibitemShut {NoStop}%
\bibitem [{\citenamefont {Miller}\ and\ \citenamefont
  {Miyake}(2015)}]{miller2015resource}%
  \BibitemOpen
  \bibfield  {author} {\bibinfo {author} {\bibfnamefont {J.}~\bibnamefont
  {Miller}}\ and\ \bibinfo {author} {\bibfnamefont {A.}~\bibnamefont
  {Miyake}},\ }\href {\doibase 10.1103/PhysRevLett.114.120506} {\bibfield
  {journal} {\bibinfo  {journal} {Phys. Rev. Lett.}\ }\textbf {\bibinfo
  {volume} {114}},\ \bibinfo {pages} {120506} (\bibinfo {year}
  {2015})}\BibitemShut {NoStop}%
\bibitem [{\citenamefont {Wilde}(2013)}]{wilde2013quantum}%
  \BibitemOpen
  \bibfield  {author} {\bibinfo {author} {\bibfnamefont {M.~M.}\ \bibnamefont
  {Wilde}},\ }\href@noop {} {\emph {\bibinfo {title} {Quantum information
  theory}}}\ (\bibinfo  {publisher} {Cambridge University Press},\ \bibinfo
  {year} {2013})\BibitemShut {NoStop}%
\bibitem [{\citenamefont {Bennett}\ \emph {et~al.}(1996)\citenamefont
  {Bennett}, \citenamefont {Bernstein}, \citenamefont {Popescu},\ and\
  \citenamefont {Schumacher}}]{bennett1996concentrating}%
  \BibitemOpen
  \bibfield  {author} {\bibinfo {author} {\bibfnamefont {C.~H.}\ \bibnamefont
  {Bennett}}, \bibinfo {author} {\bibfnamefont {H.~J.}\ \bibnamefont
  {Bernstein}}, \bibinfo {author} {\bibfnamefont {S.}~\bibnamefont {Popescu}},
  \ and\ \bibinfo {author} {\bibfnamefont {B.}~\bibnamefont {Schumacher}},\
  }\href@noop {} {\bibfield  {journal} {\bibinfo  {journal} {Phys. Rev. A}\
  }\textbf {\bibinfo {volume} {53}},\ \bibinfo {pages} {2046} (\bibinfo {year}
  {1996})}\BibitemShut {NoStop}%
\bibitem [{\citenamefont {Schollw{\"o}ck}(2011)}]{schollwock}%
  \BibitemOpen
  \bibfield  {author} {\bibinfo {author} {\bibfnamefont {U.}~\bibnamefont
  {Schollw{\"o}ck}},\ }\href@noop {} {\bibfield  {journal} {\bibinfo  {journal}
  {Annals of Physics}\ }\textbf {\bibinfo {volume} {326}},\ \bibinfo {pages}
  {96} (\bibinfo {year} {2011})}\BibitemShut {NoStop}%
\bibitem [{\citenamefont {Briegel}\ and\ \citenamefont
  {Raussendorf}(2001)}]{briegel2001persistent}%
  \BibitemOpen
  \bibfield  {author} {\bibinfo {author} {\bibfnamefont {H.~J.}\ \bibnamefont
  {Briegel}}\ and\ \bibinfo {author} {\bibfnamefont {R.}~\bibnamefont
  {Raussendorf}},\ }\href@noop {} {\bibfield  {journal} {\bibinfo  {journal}
  {Phys. Rev. Lett.}\ }\textbf {\bibinfo {volume} {86}},\ \bibinfo {pages}
  {910} (\bibinfo {year} {2001})}\BibitemShut {NoStop}%
\bibitem [{\citenamefont {Son}\ \emph {et~al.}(2012)\citenamefont {Son},
  \citenamefont {Amico},\ and\ \citenamefont {Vedral}}]{son2012topological}%
  \BibitemOpen
  \bibfield  {author} {\bibinfo {author} {\bibfnamefont {W.}~\bibnamefont
  {Son}}, \bibinfo {author} {\bibfnamefont {L.}~\bibnamefont {Amico}}, \ and\
  \bibinfo {author} {\bibfnamefont {V.}~\bibnamefont {Vedral}},\ }\href@noop {}
  {\bibfield  {journal} {\bibinfo  {journal} {Quantum Information Processing}\
  }\textbf {\bibinfo {volume} {11}},\ \bibinfo {pages} {1961} (\bibinfo {year}
  {2012})}\BibitemShut {NoStop}%
\bibitem [{\citenamefont {Else}\ \emph
  {et~al.}(2012{\natexlab{b}})\citenamefont {Else}, \citenamefont {Bartlett},\
  and\ \citenamefont {Doherty}}]{else2012symmetry}%
  \BibitemOpen
  \bibfield  {author} {\bibinfo {author} {\bibfnamefont {D.~V.}\ \bibnamefont
  {Else}}, \bibinfo {author} {\bibfnamefont {S.~D.}\ \bibnamefont {Bartlett}},
  \ and\ \bibinfo {author} {\bibfnamefont {A.~C.}\ \bibnamefont {Doherty}},\
  }\href@noop {} {\bibfield  {journal} {\bibinfo  {journal} {New Journal of
  Physics}\ }\textbf {\bibinfo {volume} {14}},\ \bibinfo {pages} {113016}
  (\bibinfo {year} {2012}{\natexlab{b}})}\BibitemShut {NoStop}%
\bibitem [{\citenamefont {Raussendorf}\ \emph {et~al.}(2019)\citenamefont
  {Raussendorf}, \citenamefont {Okay}, \citenamefont {Wang}, \citenamefont
  {Stephen},\ and\ \citenamefont {Nautrup}}]{raussendorf2019computationally}%
  \BibitemOpen
  \bibfield  {author} {\bibinfo {author} {\bibfnamefont {R.}~\bibnamefont
  {Raussendorf}}, \bibinfo {author} {\bibfnamefont {C.}~\bibnamefont {Okay}},
  \bibinfo {author} {\bibfnamefont {D.-S.}\ \bibnamefont {Wang}}, \bibinfo
  {author} {\bibfnamefont {D.~T.}\ \bibnamefont {Stephen}}, \ and\ \bibinfo
  {author} {\bibfnamefont {H.~P.}\ \bibnamefont {Nautrup}},\ }\href@noop {}
  {\bibfield  {journal} {\bibinfo  {journal} {Phys. Rev. Lett.}\ }\textbf
  {\bibinfo {volume} {122}},\ \bibinfo {pages} {090501} (\bibinfo {year}
  {2019})}\BibitemShut {NoStop}%
\bibitem [{\citenamefont {Devakul}\ and\ \citenamefont
  {Williamson}(2018)}]{devakul2018universal}%
  \BibitemOpen
  \bibfield  {author} {\bibinfo {author} {\bibfnamefont {T.}~\bibnamefont
  {Devakul}}\ and\ \bibinfo {author} {\bibfnamefont {D.~J.}\ \bibnamefont
  {Williamson}},\ }\href {\doibase 10.1103/PhysRevA.98.022332} {\bibfield
  {journal} {\bibinfo  {journal} {Phys. Rev. A}\ }\textbf {\bibinfo {volume}
  {98}},\ \bibinfo {pages} {022332} (\bibinfo {year} {2018})}\BibitemShut
  {NoStop}%
\bibitem [{\citenamefont {Daniel}\ \emph {et~al.}(2020)\citenamefont {Daniel},
  \citenamefont {Alexander},\ and\ \citenamefont
  {Miyake}}]{Daniel2020computational}%
  \BibitemOpen
  \bibfield  {author} {\bibinfo {author} {\bibfnamefont {A.~K.}\ \bibnamefont
  {Daniel}}, \bibinfo {author} {\bibfnamefont {R.~N.}\ \bibnamefont
  {Alexander}}, \ and\ \bibinfo {author} {\bibfnamefont {A.}~\bibnamefont
  {Miyake}},\ }\href {\doibase 10.22331/q-2020-02-10-228} {\bibfield  {journal}
  {\bibinfo  {journal} {{Quantum}}\ }\textbf {\bibinfo {volume} {4}},\ \bibinfo
  {pages} {228} (\bibinfo {year} {2020})}\BibitemShut {NoStop}%
\bibitem [{\citenamefont {Stephen}(2017)}]{Stephen_2017}%
  \BibitemOpen
  \bibfield  {author} {\bibinfo {author} {\bibfnamefont {D.~T.}\ \bibnamefont
  {Stephen}},\ }\emph {\bibinfo {title} {Computational power of one-dimensional
  symmetry-protected topological phases}},\ \href {\doibase
  http://dx.doi.org/10.14288/1.0354465} {Master's thesis},\ \bibinfo  {school}
  {University of British Columbia} (\bibinfo {year} {2017})\BibitemShut
  {NoStop}%
\bibitem [{\citenamefont {Verstraete}\ \emph {et~al.}(2004)\citenamefont
  {Verstraete}, \citenamefont {Popp},\ and\ \citenamefont
  {Cirac}}]{verstraete2004entanglement}%
  \BibitemOpen
  \bibfield  {author} {\bibinfo {author} {\bibfnamefont {F.}~\bibnamefont
  {Verstraete}}, \bibinfo {author} {\bibfnamefont {M.}~\bibnamefont {Popp}}, \
  and\ \bibinfo {author} {\bibfnamefont {J.~I.}\ \bibnamefont {Cirac}},\ }\href
  {\doibase 10.1103/PhysRevLett.92.027901} {\bibfield  {journal} {\bibinfo
  {journal} {Phys. Rev. Lett.}\ }\textbf {\bibinfo {volume} {92}},\ \bibinfo
  {pages} {027901} (\bibinfo {year} {2004})}\BibitemShut {NoStop}%
\bibitem [{\citenamefont {Choo}\ \emph {et~al.}(2018)\citenamefont {Choo},
  \citenamefont {Von~Keyserlingk}, \citenamefont {Regnault},\ and\
  \citenamefont {Neupert}}]{choo2018measurement}%
  \BibitemOpen
  \bibfield  {author} {\bibinfo {author} {\bibfnamefont {K.}~\bibnamefont
  {Choo}}, \bibinfo {author} {\bibfnamefont {C.~W.}\ \bibnamefont
  {Von~Keyserlingk}}, \bibinfo {author} {\bibfnamefont {N.}~\bibnamefont
  {Regnault}}, \ and\ \bibinfo {author} {\bibfnamefont {T.}~\bibnamefont
  {Neupert}},\ }\href@noop {} {\bibfield  {journal} {\bibinfo  {journal} {Phys.
  Rev. Lett.}\ }\textbf {\bibinfo {volume} {121}},\ \bibinfo {pages} {086808}
  (\bibinfo {year} {2018})}\BibitemShut {NoStop}%
\bibitem [{\citenamefont {DiVincenzo}\ \emph {et~al.}(2002)\citenamefont
  {DiVincenzo}, \citenamefont {Leung},\ and\ \citenamefont
  {Terhal}}]{divincenzo2002quantum}%
  \BibitemOpen
  \bibfield  {author} {\bibinfo {author} {\bibfnamefont {D.~P.}\ \bibnamefont
  {DiVincenzo}}, \bibinfo {author} {\bibfnamefont {D.~W.}\ \bibnamefont
  {Leung}}, \ and\ \bibinfo {author} {\bibfnamefont {B.~M.}\ \bibnamefont
  {Terhal}},\ }\href@noop {} {\bibfield  {journal} {\bibinfo  {journal} {IEEE
  Transactions on Information Theory}\ }\textbf {\bibinfo {volume} {48}},\
  \bibinfo {pages} {580} (\bibinfo {year} {2002})}\BibitemShut {NoStop}%
\end{thebibliography}%

\end{document}